\documentstyle[preprint,aps]{revtex}
\tighten
\draft
\begin{document}
\widetext
\input epsf
\preprint{CLNS 96/1433, HUTP-96/A048, NUB 3145}
\bigskip
\bigskip
\title{A Classification of $3$-Family Grand Unification in String Theory\\ 
I. The $SO(10)$ and $E_6$ Models}
\medskip
\author{Zurab Kakushadze$^1$\footnote{E-mail: 
zurab@string.harvard.edu} and S.-H. Henry Tye$^2$\footnote{E-mail: 
tye@hepth.cornell.edu}}

\medskip

\address{$^1$Lyman Laboratory of Physics, Harvard University, Cambridge, 
MA 02138 and\\
Department of Physics, Northeastern University, Boston, MA 02115\\
$^2$Newman Laboratory of Nuclear Studies, Cornell University,
Ithaca, NY 14853-5001}
\date{October 15, 1996}

\bigskip
\maketitle

\begin{abstract}

{}We give a classification of $3$-family $SO(10)$ and $E_6$ grand
unification in string theory  
within the framework of conformal (free) field theory and asymmetric 
orbifolds. We argue that the construction of such models
in the heterotic string theory requires certain ${\bf Z}_6$ asymmetric 
orbifolds that include a ${\bf Z}_3$ outer-automorphism,
the latter yielding a level-3 current algebra for the grand unification 
gauge group $SO(10)$ or $E_6$. 
We then classify all such ${\bf Z}_6$ 
asymmetric orbifolds that result in models with a non-abelian hidden sector.
All models classified in this paper have only one adjoint (but no other higher 
representation) Higgs field in the grand unified gauge group. 
In addition, all of them are completely anomaly free.
There are two types of such $3$-family models. 
The first type consists of the unique $SO(10)$ model with 
$SU(2) \otimes SU(2) \otimes SU(2)$ as its hidden sector (which
is not asymptotically-free at the string scale).
This $SO(10)$ model has $4$ left-handed and $1$ right-handed ${\bf 16}$s.
The second type is described by a moduli space containing
17 models (distinguished by their massless spectra).
All these models have an $SU(2)$ hidden sector, and $5$ left-handed and 
$2$ right-handed families in the grand unified gauge group.
One of these models is the unique $E_6$ model with an asymptotically-free 
$SU(2)$ hidden sector. The others are $SO(10)$ models, 8 of them with an 
asymptotically free hidden sector at the string scale.

\end{abstract}
\narrowtext

\section{Introduction}

\bigskip

{}If superstring theory is relevant to nature, it must contain the
standard model of strong and electroweak interactions as part of its low
energy ({\em i.e.}, below the string scale) effective field theory \cite{rev}.
There are a number of possibilities \cite{other} of embedding
the standard model in superstring.
One elegant way of realizing such an embedding is via a grand 
unified theory(GUT). Since GUT is a truly
unified theory with one gauge coupling in the low energy effective theory,
one may argue that it is the most appealing possibility.
Various GUTs in string theory have been extensively studied \cite{two}.
Since nature seems to have only three light families of quarks and
leptons, it is reasonable, and perhaps even desirable, to construct 
superstring models that contain $3$-family GUTs.
In earlier papers \cite{three,kt,five}, we have considered explicit 
realization of such $3$-family GUTs in the heterotic string theory. 
The number of such possibilities turns out to be quite limited.
In this paper, we present a classification of $3$-family $SO(10)$ and 
$E_6$ grand unification heterotic string models, within the framework
of perturbative string theory, as described below. 
The $SU(5)$ and the $SU(6)$ cases will be described in a separate paper.

{}For the grand unified gauge group to break spontaneously to the 
standard model in the low energy effective field theory, we need an adjoint 
(or other appropriate representation) Higgs field in the massless
spectrum. 
In string theory, it is most natural to have space-time supersymmetry,
which we will also impose. Since nature is not explicitly supersymmetric,
supersymmetry must be broken. The generally accepted way of achieving this is
via dynamical supersymmetry breaking in the hidden sector, which is then
transmitted to the observable sector via interactions with gravity or other
messenger/intermediate sector \cite{gaugino}. 
Dynamical supersymmetry breaking via
gaugino condensate requires an asymptotically-free hidden sector.
Motivated by these considerations, we will impose the following 
constraints on grand unified string model-building:\\
({\em i}) $N=1$ space-time supersymmetry;\\
({\em ii}) Three chiral families of fermions in the GUT gauge group;\\
({\em iii}) Adjoint (and/or other appropriate) Higgs fields in GUT;\\
({\em iv}) Non-abelian hidden sector. \\
Imposing these four constraints on model-building, within 
the framework of conformal field theory and asymmetric orbifolds,
we argue that we need 
only consider ${\bf Z}_6$ asymmetric orbifolds that include a
${\bf Z}_3$ outer-automorphism. So the classification problem is
reduced to the classification of such ${\bf Z}_6$ asymmetric orbifold models
that have the above properties.
This is carried out in this paper. It turns out that all the 
$3$-family $SO(10)$ and $E_6$ models have the following properties: \\
$\bullet$ one adjoint (plus some lower representation) Higgs 
field(s) in the GUT group; \\
$\bullet$ an intermediate/horizontal gauge symmetry.

{}This classification yields two types of models: \\
({\em i}) A unique $SU(2)_1^3 \otimes SO(10)_3 \otimes U(1)^4$ model, 
where the subscripts indicate the levels of the respective current algebras. 
This model has $4$ left-handed and $1$ right-handed ${\bf 16}$s,
{\em i.e.}, 3 chiral families, and one Higgs superfield.
The $SU(2)^3$ form its hidden sector. However, none of these three $SU(2)$s 
is asymptotically-free at the string scale. \\
({\em ii}) A unique $SU(2)_1 \otimes (E_6)_3 \otimes U(1)^3$ model, 
which lies in the same moduli space with a set of other $SO(10)$ models. 
In this sense, we should consider this as one model. 
All the models have $5$ left-handed and $2$ right-handed families.
Some of the points in this moduli space can be reached directly via
the asymmetric orbifold construction. The set of $SO(10)$ models 
obtainable this way consists of 4 subsets: \\
$\bullet$ 4 models with $SU(2)_1 \otimes SU(2)_3 \otimes SO(10)_3 \otimes 
U(1)^3$; \\
$\bullet$ 4 models with $SU(2)_1 \otimes SO(10)_3 \otimes U(1)^4$; \\
$\bullet$ 4 models with $SU(2)_1 \otimes SO(10)_3 \otimes U(1)^3$; \\
$\bullet$ 4 models with $SU(2)_1 \otimes SO(10)_3 \otimes U(1)^2$. \\
Although the 4 models in each subset have the same gauge symmetry, they 
are distinguished from each other by their massless spectra.
All the massless particles with $SO(10)$ quantum numbers are singlets under 
$SU(2)_1$, {\em i.e.}, this $SU(2)_1$ plays the role of a hidden sector.
In each of these 4 subsets, this $SU(2)_1$ in 2 of the 4 models is 
asymptotically-free, while it is not asymptotically-free in the other 2, 
at least at the string scale.
(The $SU(2)_1$ hidden sector of the $E_6$ model is also asymptotically-free.)
The $SU(2)_3$ in the first subset is not asymptotically-free at the string 
scale, and it plays the role of a horizontal symmetry. 
It also plays the role of an intermediate/messenger/mediator sector.

{}Each model in the third subset can be obtained from the corresponding model 
in either the first or the second subset via appropriate spontaneous 
symmetry breaking. Similarly, each model in the fourth subset can be 
obtained from the corresponding model in the third subset via 
appropriate spontaneous symmetry breaking. 
In this sense, one may prefer not to count
the third and the fourth subsets of models as independent models.
Since these are flat directions, it is not hard
to see that they are all connected in the same moduli space.
The $E_6$ and some of these $SO(10)$
models were constructed earlier \cite{three,kt}.

{}Note that, at the string scale, this hidden $SU(2)_1$ coupling $\alpha_2$
is three times that of the grand unified gauge group. Using the
accepted grand unified gauge coupling extracted from experiments,
the $SU(2)_1$ coupling is expected to become strong at a scale above the 
electroweak scale.
Naively, gaugino condensation in the hidden sector will not stabilize 
the dilaton expectation value to some finite reasonable value \cite{ds}, 
implying that these models may not be phenomenologically viable. 
However, recent ideas on the K\"ahler potential actually suggest 
otherwise \cite{bd}. So the issue remains open. Clearly a better
understanding of the string dynamics will be most important.

{}The rest of this paper is organized as follows. 
In section II we discuss the the general grounds underlying the 
classification of models.
Next, we turn to the construction of the models, which 
involves two steps. First, the appropriate $N=4$ supersymmetric
models, namely the Narain models, are constructed in Section III.
Then, appropriate asymmetric ${\bf Z}_6$ orbifolds of the Narain 
models that yield the various $3$-family string GUT models are 
classified in section IV. 
The massless spectra of the resulting $N=1$ models may be worked out using
the approach given in Ref \cite{kt}. Appendix A gives a brief introduction
to the construction. The models and their massless
spectra are listed in the collection of Tables. 
In section V we discuss the moduli space of
the models and the ways they are connected to each other. 
This is done in two steps to make the discussion easier to follow. 
First, we describe these connections in terms of the flat directions 
of the scalar fields in effective field theory. 
Then we do the same in terms of the string moduli, and translate 
the field theory discussion into the stringy language. 
Finally, in section VI we conclude with some remarks. Appendix B contains 
some details.

\section{Preliminaries}

\bigskip

{}Before the classification and the explicit construction of 
$3$-family GUT string models, 
we begin by outlining the general grounds under the classification.

({\em i}) Phenomenologically, the gauge coupling at the grand unification 
scale is weak. 
Since this scale is quite close to the string scale, the coupling
at the string scale is also expected to be weak. This means that string model
construction is governed by the underlying conformal field theory.
So model-building can be restricted to the four-dimensional heterotic
string theory using conformal field theories and current
(or Kac-Moody) algebras. Our framework throughout will be 
the asymmetric orbifold construction \cite{orb}. 
In working out the spectra of such 
models based on asymmetric orbifolds, the rules of Ref.\cite{kt} 
are very useful.

({\em ii}) By the three chiral families constraint,
we mean that the net number of chiral families (defined as the number of
left-handed families minus the number of the right-handed families) must be
three. The remaining pairs may be considered to be Higgs superfields. 
Of course, there is no guarantee that any
right-handed families present will always pair up with left-handed
families to become heavy. This depends on the details of dynamics. 

({\em iii}) As we mentioned earlier, the hidden sector is required so that 
dynamical supersymmetry breaking may occur
(although it still depends on the details of string dynamics 
whether supersymmetry is broken or not).
This means the hidden sector must become strongly interacting at some scale
above the electroweak scale, which in turn implies that the hidden sector
must be asymptotically-free. Here we only impose the condition that the
hidden sector must be non-abelian. This is not quite as strong as the
asymptotically-free condition, since, with enough matter fields, a
non-abelian gauge symmetry need not be asymptotically-free. However,
depending on the details of the string
dynamics, it is possible that, although not asymptotically-free at the
string scale, the hidden sector can become asymptotically-free below some
lower scale where some of its matter fields acquire masses.

({\em iv}) Let us now turn to the requirement of adjoint Higgs in GUT.
In the low energy effective theory with a grand unification gauge
symmetry, it is well known that an adjoint (or other appropriate
higher dimensional representation) Higgs field is needed to break the 
grand unified gauge group to the standard model. 
It is relatively easy to realize gauge symmetries
with level-$1$ current algebras in string theory. 
Unfortunately, such models cannot 
have $N=1$ supersymmetry, chiral fermions and adjoint (or higher dimensional) 
Higgs field all at the same time. 
This is due to the following. Level-1 current algebras do not have 
adjoint (or higher dimensional) irreducible representations (irreps) 
coming from its conformal highest weight since the latter are not 
compatible with unitarity. Thus, the only way
an adjoint Higgs can appear in the massless spectrum is that it 
comes from the identity irrep (the first descendent of the identity weight 
is the adjoint, other higher irreps are massive), the same way the 
gauge supermultiplet 
(which also transforms in the adjoint) appears in the massless spectrum. 
Since there is no distinction between the gauge quantum numbers of the 
gauge and adjoint Higgs supermultiplets, they ultimately combine into 
an $N=2$ gauge supermultiplet, and all the fields transforming in the 
irreps of this gauge group 
turn out to have $N=2$ global supersymmetry. 
This necessarily implies that the fermions are non-chiral.
So one cannot construct a string model with both chiral
fermions and adjoint Higgs using level-$1$ current algebras.
For the same reason of unitarity (as mentioned above), massless
Higgs in irreps higher than the adjoint cannot appear in level-1 models.

{}The above discussion leads one to conclude that, in order 
to incorporate both chiral fermions and adjoint 
(or higher dimensional representation) Higgs fields, 
the grand unified gauge symmetry must be realized via 
a higher-level current algebra.
Indeed, adjoint (and some higher dimensional) irreps are allowed 
by unitarity in these higher-level realizations.
This means that one can hope to construct models with adjoint Higgs which 
comes from the adjoint irrep rather than from the identity irrep.
Since now the adjoint Higgs and gauge supermultiplet come from two 
different irreps of the current algebra, they no longer combine into an $N=2$
gauge supermultiplet, and chiral fermions become possible. 
One could also hope to construct models with Higgs fields in 
higher dimensional irreps.

{}Level-2 string models, {\em i.e.}, string model with the
grand unified gauge symmetry realized via a level-2 current algebra,
have been extensively explored in the literature \cite{two}.
So far, all the known level-2 models have an even number of chiral families.
Although there is no formal proof that three-family models
based on level-$2$ current algebras do not exist, there is a simple way to
understand the failure to construct them, at least in the orbifold framework.
Level-2 models typically require ${\bf Z}_2$ (outer-automorphism)
orbifolds. The number of fixed points of such an orbifold is always
even, typically powers of 2.
Since this number determines the number of chiral families, one ends up
with an even number of chiral families. One possible way to obtain $3$ 
families is to have three different twisted sectors where each contributes
only one ({\em i.e.}, 2 to the zeroth power) family. In fact, this is the
approach used to construct interesting $3$-family standard models \cite{alon}.
However, rather exhaustive searches\cite{two} seem to rule out this 
possibility for GUT string models: there is simply not enough room to 
incorporate these different sectors, simply
because the level-$2$ GUT group takes up more room than the standard model.

{}Thus, to achieve $3$ chiral families, it is natural to go to
level-3 models, since their construction typically requires a ${\bf Z}_3$
outer-automorphism. This can be part of a ${\bf Z}_3$ orbifold, which has an
odd number of fixed points. So there is a chance to construct models
with three chiral families via ${\bf Z}_3$ orbifolds.
Since ${\bf Z}_4$ orbifolds involve an even
number of fixed points, level-4 models are likely to share the same fate
of their level-2 counterparts.
Now, higher level current algebras have larger central charges. Since
the total central charge is 22, a string model with higher-level current
algebras will have less room for the hidden sector. In the level-$3$ 
$SO(10)$ or $E_6$ models
presented in this paper, a typical hidden sector has an $SU(2)$ as its
non-abelian gauge symmetry. This indicates that level-4 and higher level 
models will not have any room for a non-abelian hidden sector. 
In fact, it is not at all clear that level-4 models can even be constructed.
Thus, level-$3$ models seem to be the only possibility.

{}Naively, one might expect that three-family string GUTs can be 
constructed with a single ${\bf Z}_3$ twist, since the latter can,
at least in principle, be arranged to have three fixed points. 
This, however, turns out not to be the case.
It can be shown (by considering all the possible ${\bf Z}_3$ twists
compatible with the following requirements) that if a model with
$SO(10)$ or $E_6$ gauge group realized at level-3 is constructed
by a single ${\bf Z}_3$ orbifold of an $N=4$ Narain lattice,
then one of the following happens (see Appendix B for more details):\\
$\bullet$ The model has $N=1$ SUSY, but no chiral fermions;\\
$\bullet$ The model has $N=1$ SUSY, and chiral fermions; then the
number of chiral families is always 9, with no
anti-generations ({\em i.e.}, 9 left-handed families
and no right-handed families);\\
$\bullet$ The model has $N=2$ SUSY, and thus no chiral fermions.

{}Since the first case is already a non-chiral $N=1$ model to start with, 
its ${\bf Z}_2$ orbifold cannot yield an odd number of chiral families (see Appendix B for details).
This implies that, to obtain a model with three chiral families, we must
start from one of the last two cases just described and
${\bf Z}_2$ orbifold it.
These are the only possibilities since ${\bf Z}_2$ is the only other 
symmetry that a Narain $\Gamma^{6,22}$ lattice\cite{narain} can
have and at the same time satisfy other requirements like, say,
$SO(10)^3$ gauge symmetry. Here we should mention that the resulting
orbifold must be ${\bf Z}_6 ={\bf Z}_3 \otimes {\bf Z}_2$, and it
cannot be a non-abelian discrete group such as, say, $S_3$.
This is because there is only one embedding of $SO(10)_3$ and
$(E_6)_3$ current algebras (with the subscripts indicating the 
levels of the current algebras) in the Narain lattice, namely, the
diagonal embedding \cite{dien}, where $SO(10)_3$ and $(E_6)_3$ groups are the diagonal
subgroups of $SO(10)^3$ and $(E_6)^3$.
The diagonal subgroup is invariant under the outer-automorphism
of the three $SO(10)$s or the three $(E_6)$s. 
This outer-automorphism is isomorphic to a
${\bf Z}_3$ twist. Thus, if the subsequent ${\bf Z}_2$ twist does not
commute with the ${\bf Z}_3$ twist, the gauge group will be further
broken, and the rank will be reduced. This is why the
${\bf Z}_2$ twist must commute with the ${\bf Z}_3$ twist.
This leads us to a classification of all ${\bf Z}_6$ asymmetric 
orbifolds that lead to GUT models with three chiral families of fermions.
Since it has been shown\cite{dien} that only diagonal embedding is 
available for $SO(10)_3$ and $E_6$, ${\bf Z}_3$ orbifolds will be able 
to reach all such models.

{}The above discussion reduces the classification problem to the
classification of ${\bf Z}_6$ asymmetric orbifold models with
the following phenomenological requirements
(that translate into stringent constraints in the actual string
model building) that we will impose: \\
$\bullet$ $N=1$ space-time supersymmetry;\\
$\bullet$ Level-3 GUT;\\
$\bullet$ Three chiral families in GUT;\\
$\bullet$ Non-abelian hidden sector.\\
Imposing these minimal phenomenological requirements, we find the unique 
$E_6$ model and its accompanying $SO(10)$ models, and the single $SO(10)$ 
model. This classification automatically 
includes the 3-family models constructed earlier \cite{three,kt}.
All of the models are given in this paper and their
massless spectra are presented in the Tables.
In Ref \cite{kt}, details of the construction are provided. Appendix A 
gives a brief outline. Since
the constructions of all the models are very similar, our discussion
here will be quite brief.

{}Now the classification project consists of three steps: \\
({\em i}) list all suitable $N=4$ Narain models \cite{narain}; \\
({\em ii}) list all ${\bf Z}_6$ asymmetric twists/shifts that can act on 
each Narain model; \\
({\em iii}) work out the massless spectrum in each orbifold. \\
Now, it is well-known that some orbifolds, though look different at first 
sight, nevertheless yield identical models.
There are two situations to be considered: \\
$\bullet$ Their spectra in each twisted/untwisted sector are identical.
Examples of such equivalent orbifolds have been discussed in some detail
in Ref \cite{kt}. In this paper, we shall not discuss this situation 
further, but simply consider one choice in each set of equivalent
${\bf Z}_6$ asymmetric twists/shifts. \\ 
$\bullet$ Their massless spectra are identical, but some particular massless
particles may appear in different sectors in different orbifolds.
Since this situation is not discussed in Ref \cite{kt}, we shall give 
the equivalent ${\bf Z}_6$ twists/shifts in this paper. These models are 
simply related by $T$-duality.

{}We should point out that the classification only distinguishes different
string models by their tree-level massless spectra and/or interactions.
For example, a typical model will have some moduli that can be varied 
(within some ranges) without affecting its massless spectrum.
Here we shall not distinguish such models. We will, however, give a 
rather detailed discussion of the moduli space in which these models sit.
This explains our emphasis on the enhanced symmetry points in the moduli space.
If we move away from those special points in the moduli space, 
one of the following happens:\\
$\bullet$ The resulting model has the same tree-level massless spectrum 
but the couplings are continuously varied: \\ 
$\bullet$ The resulting model exists, but part of its gauge 
symmetry is spontaneously broken: in particular, it may happen that
the hidden sector is broken.

{}Let us summarize some of the key features of these models.
Each model has a gauge symmetry $G \otimes M \otimes H$,
where the observable sector $G$ is either $SO(10)_3$ or $(E_6)_3$.
$H$ is the hidden sector gauge symmetry: massless $G$ supermultiplets
are by definition singlets in $H$. The remaining gauge symmetry is
referred to as $M$. Both massless $G$ supermultiplets and 
massless $H$ supermultiplets can have non-zero quantum numbers in $M$.
If massless $H$ supermultiplets are singlets under certain part of $M$, 
that part of $M$ may be considered as a horizontal symmetry. 
The rest of $M$ may be considered as the gauge group for the
messenger/mediator sector, linking the hidden and the visible sectors.
In our classification, $M$ is always composed of some factors of 
$U(1)$s and sometimes a (non-asymptotically-free) $SU(2)$.
In contrast to some of the $SU(5)$ models \cite{five},
all the $U(1)$s in the
$SO(10)$ and $E_6$ models are anomaly-free. 
Also note that some of the models do not 
have gauge singlets completely neutral under all gauge symmetries.

{}Each model has only one adjoint (but no higher representation) 
Higgs field in the grand unified gauge group.
They also have Higgs fields in lower representations.
There are two types of $SO(10)$ models. 
There is one unique $SO(10)$ model with
$SU(2)_1 \otimes SU(2)_1 \otimes SU(2)_1$ as 
its hidden sector, but none of these three $SU(2)$s is 
asymptotically-free at the string scale.
This $SO(10)$ model has $4$ left-handed and $1$ right-handed ${\bf 16}$s,
{\em i.e.}, 3 chiral families and 1 Higgs superfield in the spinor 
representation. In all the $SO(10)$ models, each ${\bf 16}$ (left or 
right-handed) is accompanied by a ${\bf 10}$ and a ${\bf 1}$. 
The other type consists of the unique $E_6$ model and a set of $SO(10)$
models. The hidden sector in the unique $E_6$ model and
its accompanying $SO(10)$ models 
is $SU(2)_1$, which is asymptotically-free at the string scale for 
some models, but not for others.
The $E_6$ model has $5$ left-handed and $2$ right-handed ${\bf 27}$s.
These $SO(10)$ models have $5$ left-handed and $2$ right-handed
${\bf 16}$s. This means that they have 3 chiral families and two Higgs 
superfields in the fundamental or spinor representation. 

{}Ignoring string threshold effects \cite{kap},
the gauge coupling of a given group $G$ in the model at a
scale $\mu$ below the string scale $M_s$ is related to it via:
\begin{equation}
 1/\alpha_G(\mu)= k_G/\alpha_{\mbox{string}}
 + ({b_0/{4\pi}}) \ln ({M_s^2/ \mu^2})~,
\end{equation}
where $k_G$ is the level of the gauge group. 
For $U(1)$ gauge symmetries, let us introduce the following convention. 
Demanding that all particles have integer $U(1)$ charges,
the level of a $U(1)$ gauge symmetry is given by $1/k = 2r^2$, where $r$
is the compactification radius of the corresponding left-moving world-sheet
boson. The $U(1)$ charge of a particle with charge $n$ contributes 
$n^2r^2/2$ to its conformal highest weight.
(Alternatively, if we choose to define the level of all $U(1)$ gauge
symmetry to be $k=1$, then that $U(1)$ charge of the particle will be 
${\sqrt 2} nr$.)
The constant $b_0$ is the one-loop coefficient of the beta-function.
Note that at the string scale, the hidden $SU(2)_1$ coupling $\alpha_2$ 
is three times that of the grand unified gauge group. 

{}Now, $SU(5)_3$ gauge symmetry
may be obtained via a diagonal embedding in $SU(5)^3$ as well as
a non-diagonal embedding \cite{dien} in $SU(10)_1$; so its
classification requires additional consideration. 
Furthermore, $SU(5)$ models allow a hidden sector with an
asymptotically-free semi-simple gauge group at the string scale;
this important phenomenological feature is absent in the
$SO(10)$ and $E_6$ models. As we have seen in
Ref\cite{five}, $SU(6)$ models are closely related to the $SU(5)$
construction, so their classification should go together.
The classification of $SU(5)$ and
$SU(6)$ models with 3 chiral families will be presented separately.

\section{Wilson Lines and Narain Models}

\bigskip

{}Starting from a consistent heterotic string model, typically
a Narain model, a new model can be generated by performing a consistent
set of twists and shifts (of the momentum lattices) on it.
First, these twists and shifts must be consistent with the lattice symmetry.
To obtain a new consistent string model, the following conditions are
imposed on its one-loop partition function in the light-cone gauge: \\
($\em i$) one-loop modular invariance; \\
($\em ii$) world-sheet supersymmetry, which insures space-time Lorentz
invariance in the covariant gauge; and \\
($\em iii$) the physically sensible projection; this means the contribution
of a space-time fermionic (bosonic) degree of freedom to the partition
function counts as minus (plus) one. In all cases that can be checked, this
condition plus the one-loop modular invariance and factorization imply
multi-loop modular invariance.

{}A set of twists and shifts can be organized into a set of vectors.
The rules of consistent asymmetric orbifold model construction can then 
be written as constraints on these vectors, as done in Ref\cite{kt}.
All sets of vectors given in this and the next sections,
${\em i.e.}$, the Wilson lines $(U_1, U_2)$, and the ${\bf Z}_6$ orbifolds
$(T_3, T_2)$, have been checked to satisfy
these consistency constraints. The spectrum of the resulting model can be
obtained from the spectrum generating formula.
A brief outline can be found in Appendix A.

{}Before we list the twists and shifts that generate the orbifold models,
we have to construct the Narain lattices that we are going to orbifold. 
This is done in two steps. First, we describe the $N=4$ Narain model 
that is obtained by compactifying the ${\mbox{Spin}}(32)/{\bf Z}_2$ 
heterotic string theory on a six-torus. Next, we will classify the 
Wilson lines that give various Narain models which contain 
$SO(10)^3$ and $(E_6)^3$ subgroups. 
${\bf Z}_6$ orbifolds of these models will then give rise to 
the $SO(10)_3$ and $(E_6)_3$ models with three families of chiral fermions. 

\subsection{Moduli Space of $N=4$ Models}
\medskip

{}Consider the Narain model
with the momenta of the internal bosons spanning an even self-dual 
Lorentzian lattice $\Gamma^{6,22} =\Gamma^{6,6} \otimes
\Gamma^{16}$. Here $\Gamma^{16}$ is the  
${\mbox{Spin}}(32)/{\bf Z}_2$ lattice. The $\Gamma^{6,6}$ is the 
momentum lattice corresponding to the compactification on a six-torus 
defined by $X_I= X_I +E_I$. The dot product of the vectors $E_I$ 
defines the constant background metric $G_{IJ}=E_I \cdot E_J$. 
There is also the antisymmetric background field $B_{IJ}$. 
The components of $G_{IJ}$ and $B_{IJ}$ parametrize the $36$-dimensional 
moduli space of the $\Gamma^{6,6}$ lattice. 
We are only interested in the subspace of the moduli space that has 
(1) appropriate ${\bf Z}_3$ symmetries, upon which we have to perform a
${\bf Z}_3$ twist later, and (2) an enhanced gauge symmetry so that,
after the orbifold, the hidden sector can have maximal gauge symmetry.
With these constraints, it will be suffice to consider a two-dimensional 
subspace of the general  moduli space. 
This subspace is parametrized by the moduli
$h$ and $f$, and the vectors $E_I$ (and also their duals ${\tilde E}^I$
defined so that $E_I \cdot {\tilde E}^J ={\delta_I}^J$) can be expressed 
in terms of the $SU(3)$ root and weight vectors
${e}_i$ and ${\tilde e}^i$ ($i=1,2$):
\begin{eqnarray}\label{}
 &&E_1=(e_1,0,0),~~~E_2=(e_2,0,0),\nonumber\\
 &&E_3=(0,e_1,0),~~~E_4=(0,e_2,0),\nonumber\\
 &&E_5=(f{\tilde e}^2,-h {\tilde e}^2,ge_1),~~~E_6=(-f{\tilde e}^1,h {\tilde e}^1,ge_2),\nonumber 
\end{eqnarray}
where $g=\sqrt{1-(f^2+h^2)/3}$. 
The components of the antisymmetric tensor are chosen to be
$2B_{IJ} ={1\over 2} G_{IJ}$ for $I<J$ and $2B_{IJ} =-{1\over 2} G_{IJ}$ 
for $I>J$. We shall call these Narain models as $N(h,f)$.
 
{}With the above choice of the $\Gamma^{6,6}$ lattice, the Narain model $N(h,f)$ has the gauge symmetry $R(h,f) \otimes SO(32)$. Let us consider a the region $0\leq h,f\leq 1$.
A generic point in the moduli space of $N(h,f)$ has $R(h,f)=
SU(3) \otimes SU(3)\otimes U(1)^2$. 
There are four isolated points with enhanced gauge symmetry: \\
$\bullet$ $R(0,0)=SU(3) \otimes SU(3)\otimes SU(3)$, \\
$\bullet$ $R(1,0)=R(0,1)=SO(8) \otimes SU(3)$ and \\
$\bullet$ $R(1,1)=E_6$. 

{}At these points, the $\Gamma^{6,6}$ lattice can be generated by the 
Lorentzian vectors $(0 \vert\vert E_I)$ and 
$({\tilde E}^I \vert\vert {\tilde E}^I)$. 
For our purpose, we only need to consider the three isolated points, 
{\em i.e}, $(h,f)=(0,0)$, $(1,0)$ and $(1,1)$ ($(h,f)=(0,1)$
case is equivalent to the $(h,f)=(1,0)$ case), and certain deformations of the lattice that start at these isolated points.

\subsection{Wilson Lines}
\medskip

{}Next, we discuss the  Wilson lines that will give the Narain models 
with $SO(10)^3$ and $(E_6)^3$ gauge subgroups.   
Here we are writing the Wilson lines as shift vectors in the $\Gamma^{6,22}$
lattice. The shift vectors $U_1$ and $U_2$ to be introduced 
are order-$2$ shifts that break the $SO(32)$ to $SO(10)^3 \otimes SO(2)$. 
This will be the generic case, although 
sometimes, the new $U$ sectors can introduce additional gauge bosons to 
enhance $SO(10)^3$ to $(E_6)^3$. The remaining gauge symmetry
must contain a non-abelian gauge group with a ${\bf Z_3}$ symmetry.
This will allow a non-abelian gauge group to emerge after the orbifolding. From 
the above discussion, this condition restricts us to the cases with the
moduli $(h,f)=(0,0)$, $(1,0)$ and $(1,1)$.

{}Now, string consistency imposes constraints on the choice of the
set of shift vectors. In each case we consider below,
it is easy to check that the set of shift vectors ($U_1$, $U_2$) satisfies
the string consistency conditions. (See Ref \cite{kt} for more details.)
Since all the gauge symmetries in Narain models are realized with 
level-$1$ current algebras, we shall skip the level 
labelling in this section.

For $h=f=1$ we have two inequivalent choices:\\
$\bullet$ The $N1(1,1)$ model generated by the Wilson lines
\begin{eqnarray}
 &&U_1 =(0,0,0 \vert\vert e_1/2,0,0)({\bf s}\vert {\bf 0}\vert {\bf 0}\vert 
 {\overline    S})~,\nonumber\\
 &&U_2 =(0,0,0 \vert\vert e_2/2,0,0)({\bf 0}\vert {\bf s}\vert {\bf 0}\vert {\overline S})~.\nonumber
\end{eqnarray} 
This model has $SU(3)^2 \otimes (E_6)^3$ gauge symmetry.
The $U_1$ and $U_2$ are order-$2$ (${\bf Z}_2$) shifts. 
The first three entries correspond to the right-moving complex world-sheet
bosons. 
The next three entries correspond to the left-moving complex world-sheet
bosons. Together they form the six-torus. The remaining 16 left-moving 
world-sheet bosons generate the ${\mbox{Spin}}(32)/{\bf Z}_2$ lattice. 
The $SO(32)$ shifts are given in the $SO(10)^3
\otimes SO(2)$ basis. In this basis, ${\bf 0}$ stands for the null vector, 
${\bf v}$($V$) is the vector weight, whereas
${\bf s}$($S$) and ${\overline {\bf s}}$(${\overline S}$) are the
spinor and anti-spinor weights of $SO(10)$($SO(2)$). (For $SO(2)$, $V=1$,
$S=1/2$ and ${\overline S}=-1/2$.) 
The unshifted sector provides gauge bosons of
$SU(3)^{2} \otimes U(1)^{2} \otimes SO(10)^{3} \otimes SO(2)$. 
The permutation symmetry of the three $SO(10)s$ are explicit here. 
There are additional gauge bosons from the new sectors. Recall that under
$E_6 \supset SO(10) \otimes U(1)$, 
\begin{equation}
{\bf 78}={\bf 1}(0)+{\bf 45}(0)+{\bf 16}(3)+\overline{\bf 16}(-3)~.
\end{equation}
It is easy to see that the $U_1$, $U_2$ and $U_1+U_2$ sectors provide
the necessary ${\bf 16}(3)$ and $\overline{\bf 16}(-3)$ gauge
bosons to the three $SO(10)$'s respectively. Consistency and the 
permutation symmetry of the three $SO(10)$s implies the permutation 
symmetry of the three $E_6$s.
The resulting Narain $N1(1,1)$ model has $N=4$ SUSY and gauge group
$SU(3)^2 \otimes (E_6)^3$ provided that we set $k_{10}=k_{20}=0$. \\ 
$\bullet$ The $N2(1,1)$ model generated by the Wilson lines
\begin{eqnarray}
 &&U_1 =(0,e_1/2,e_1/2 \vert\vert e_1/2,0,0)({\bf s}\vert {\bf 0}\vert {\bf 0}\vert 
 {\overline S})~,\nonumber\\
 &&U_2 =(0,e_2/2,e_2/2\vert\vert e_2/2,0,0)({\bf 0}\vert {\bf s}\vert {\bf 0}\vert 
 {\overline S})~.\nonumber
\end{eqnarray}  
This model has  
$SU(3)^2 \otimes U(1)^2\otimes SO(10)^3 \otimes SO(2)$ gauge symmetry. 

{}For $h=f=0$, we have four inequivalent choices:\\
$\bullet$ The $N1(0,0)$ model  generated by the Wilson lines
\begin{eqnarray}
 &&U_1 =(e_1/2,e_1/2,e_1/2 \vert\vert 0,0,0)({\bf s}\vert {\bf 0}\vert {\bf 0}\vert 
 {\overline    S})~,\nonumber\\
 &&U_2 =(e_2/2,e_2/2,e_2/2 \vert\vert  0,0,0)({\bf 0}\vert {\bf s}\vert {\bf 0}\vert 
 {\overline S})~.\nonumber
\end{eqnarray}  
This model has  $SU(3)^3 \otimes SO(10)^3 \otimes SO(2)$ gauge symmetry.\\
$\bullet$ The $N2(0,0)$ model  generated by the Wilson lines
\begin{eqnarray}
 &&U_1 =(e_1/2,e_1/2,0 \vert\vert 0,0,e_1/2)({\bf s}\vert {\bf 0}\vert {\bf 0}\vert 
 {\overline    S})~,\nonumber\\
 &&U_2 =(e_2/2,e_2/2,0 \vert\vert  0,0,e_2/2)({\bf 0}\vert {\bf s}\vert {\bf 0}\vert 
 {\overline    S})~,\nonumber
\end{eqnarray}  
This model has $SU(3)^2\otimes U(1)^2 \otimes SO(10)^3 \otimes SO(2)$
gauge symmetry.\\
$\bullet$ The $N3(0,0)$ model  generated by the Wilson lines
\begin{eqnarray}
 &&U_1 =(e_1/2,0,0 \vert\vert 0,e_1/2,e_1/2)({\bf s}\vert {\bf 0}\vert {\bf 0}\vert 
 {\overline    S})~,\nonumber\\
 &&U_2 =(e_2/2,0,0 \vert\vert  0,e_2/2,e_1/2)({\bf 0}\vert {\bf s}\vert {\bf 0}\vert 
 {\overline    S})~,\nonumber
\end{eqnarray}  
This model has  $SU(3)\otimes U(1)^4 \otimes SO(10)^3 \otimes SO(2)$
gauge symmetry.\\
$\bullet$ The $N4(0,0)$ model  generated by the Wilson lines
\begin{eqnarray}
 &&U_1 =(e_1/2,0,0 \vert\vert 0,0,0)({\bf s}\vert {\bf 0}\vert {\bf 0}\vert 
 {\overline    S})~,\nonumber\\
 &&U_2 =(e_2/2,0,0 \vert\vert  0,0,0)({\bf 0}\vert {\bf s}\vert {\bf 0}\vert 
 {\overline    S})~,\nonumber
\end{eqnarray}  
This model has  $SU(3)^3 \otimes SO(10)^3 \otimes SO(2)$
gauge symmetry.

{}For $h=1$, $f=0$, we have two inequivalent choices: \\
$\bullet$ The $N1(1,0)$ model  generated by the Wilson lines
\begin{eqnarray}
 &&U_1 =(e_1/2,a_1,b_1\vert\vert 0,0,0)({\bf s}\vert {\bf 0}\vert {\bf 0}\vert 
 {\overline    S})~,\nonumber\\
 &&U_2 =(e_2/2,a_2,b_2 \vert\vert  0,0,0)({\bf 0}\vert {\bf s}\vert 
 {\bf 0}\vert {\overline S})~.\nonumber
\end{eqnarray}  
Here the four component vectors $(a_1,b_1)$ and $(a_1,b_1)$ are the spinor 
${\bf  s}$ and conjugate $ {\bf  c}$ weights of  $ SO(8)$, respectively.
This model has  $SU(3)\otimes SO(8)\otimes SO(10)^3 \otimes SO(2)$ gauge 
symmetry.\\
$\bullet$ The $N2(1,0)$ model  generated by the Wilson lines
\begin{eqnarray}
 &&U_1 =(e_1/2,0,0\vert\vert 0,0,0)({\bf s}\vert {\bf 0}\vert {\bf 0}\vert 
 {\overline    S})~,\nonumber\\
 &&U_2 =(e_2/2,0,0 \vert\vert  0,0,0)({\bf 0}\vert {\bf s}\vert {\bf 0}\vert 
 {\overline S})~.\nonumber
\end{eqnarray}  
This model has $SU(3)\otimes SO(8)\otimes SO(10)^3 \otimes SO(2)$ gauge 
symmetry. 

{}This completes the list of $N=4$ supersymmetric models that are
the starting points for orbifolding.

\bigskip

\section{A Classification of ${\bf Z}_6$ Asymmetric Orbifolds}
\medskip

{}Here we list all possible ${\bf Z}_6$ asymmetric orbifolds that can 
be performed on the above list of Narain models to yield $3$-family GUT 
models. To make the discussion easier to follow, we will split the
${\bf Z}_6$ twist into a ${\bf Z}_3$ twist accompanied by a ${\bf Z}_2$ twist.
The action of the ${\bf Z}_3$ twist on the states corresponding to the 
original $\Gamma^{16}$ sublattice
is fixed by the requirement that we have $SO(10)_3$ or 
$(E_6)_3$ current algebra and chiral fermions. 

{}The action of the ${\bf Z}_2$ twist on the $\Gamma^{16}$ states is also 
fixed, namely, the ${\bf Z}_2$ twist does not act on the $\Gamma^{16}$ at all. 
This point is not hard to see.
{\em A priori}, we could include a 
${\bf Z}_2$ shift in the $SO(2)$ lattice. 
This, however, would destroy the permutational symmetry of states in sectors
$U_1 \rightarrow U_2 \rightarrow U_1+U_2 \rightarrow U_1$. 
This symmetry is required for the ${\bf Z}_3$ orbifold that leads to 
the corresponding level three current algebra. In fact, no other
${\bf Z}_2$ twist or shift can act on the $SO(2)$ lattice 
(and on the $\Gamma^{16}$ states, in general) for the same reason. 
This, in particular, implies that (starting from the models classified 
in this paper) we cannot enhance the gauge symmetry in the hidden sector 
by adding a ${\bf Z}_2$ Wilson line and absorbing a $U(1)$ factor that 
has $SO(2)$ admixture
and still maintain the $SO(10)_3$ or $(E_6)_3$ gauge symmetry. 
Note, however, that if we break $SO(10)_3$ to $SU(5)_3 \otimes U(1)$ by 
a ${\bf Z}_3$ Wilson line, then the hidden sector gauge symmetry can be 
enhanced. An example of this was given in Ref\cite{five}.

{}Before we describe the  ${\bf Z}_6$ asymmetric orbifolds, 
we will introduce some notation. By $\theta$ we will denote a 
$2\pi/3$ rotation of the corresponding two real chiral world-sheet bosons. 
Thus, $\theta$ is a ${\bf Z}_3$ twist. 
Similarly, by $\sigma$ we will denote a $\pi$ rotation of the 
corresponding two chiral world-sheet bosons. 
Thus, $\sigma$ is a ${\bf Z}_2$ twist. By ${\cal P}$ we will denote
the outer-automorphism of the three $SO(10)$s that arise in the 
breaking $SO(32)\supset SO(10)^3 \otimes SO(2)$. 
Note that ${\cal P}$ is a ${\bf Z}_3$ twist. Finally, by $(p_1,p_2)$ 
we will denote the outer-automorphism of the corresponding two complex 
chiral world-sheet bosons.
Note that $(p_1,p_2)$ is a ${\bf Z}_2$ twist. The spin structures
of the world-sheet fermions in the right-moving sector are fixed by the 
world-sheet supersymmetry consistency. Again, the string consistency
conditions impose tight constraints on the allowed twists. Using the
approach given in Ref \cite{kt}, and briefly reviewed in Appendix A, it is 
quite easy to check that
each of the sets of twists introduced below are consistent, provided 
appropriate choices of the structure constants $k_{ij}$ are picked. It is then 
straightforward, but somewhat tedious, to work out the massless spectrum 
in each model. (Again, more details can be found in Ref \cite{kt}. See also
Appendix  A.)

{}Finally, we are ready to give the possible ${\bf Z}_3 \otimes {\bf Z}_2$ 
twists.\\
$\bullet$ The $E1$ model. Start from the $N1(1,1)$ model and perform 
the following twists:
\begin{eqnarray}
 &&T_3 =(\theta,\theta,\theta\vert\vert \theta,e_1/3,0)
 ({\cal P} \vert  2/3)~,\nonumber\\
 &&T_2=(\sigma, p_1,p_2\vert\vert  0,e_1/2,e_1/2)
 (0^{15} \vert 0)~.\nonumber
\end{eqnarray}  
This model has $SU(2)_1 \otimes (E_6)_3 \otimes U(1)^3$ gauge symmetry. 
The massless spectrum of the $E1$ model is given in Table I. They are 
grouped according to where they come from, namely, the untwisted sector U,
the ${\bf Z}_3$ twisted ({\em i.e.}, $T_3$ and $2T_3$) sector T3, 
the ${\bf Z}_6$ twisted ({\em i.e.}, $T_3+T_2$ and $2T_3+T_2$) sector T6,
and ${\bf Z}_2$ twisted ({\em i.e.}, $T_2$) sector T2.\\
$\bullet$ The $E2$ model. Start from the $N1(1,1)$ model and perform the 
following twists:
\begin{eqnarray}
 &&T_3 =(0,\theta,\theta\vert\vert \theta,e_1/3,0)
 ({\cal P} \vert  2/3)~,\nonumber\\
 &&T_2=(\sigma, p_1,p_2\vert\vert  0,e_1/2,e_1/2)
 (0^{15} \vert 0)~.\nonumber
\end{eqnarray}  
This model has $SU(2)_1 \otimes (E_6)_3 \otimes U(1)^3$ gauge symmetry. 
The massless spectrum of the $E2$ model is given in Table I.\\
Here we note that the $E1$ and $E2$ models are the same, in particular, 
they have the same tree-level massless spectra and interactions.\\
$\bullet$ The $T1(1,1)$ model. Start from the $N2(1,1)$ model and perform 
the same twists as in the $E1$ model.
This model has $SU(2)_1 \otimes SO(10)_3 \otimes U(1)^4$ gauge symmetry. 
The massless spectrum of the $T1(1,1)$ model is given in Table II.\\
$\bullet$ The $T2(1,1)$ model. Start from the $N2(1,1)$ model and perform the 
same twists as in the $E2$ model. 
This model has $SU(2)_1 \otimes SO(10)_3 \otimes U(1)^4$ gauge symmetry. 
The massless spectrum of the $T2(1,1)$ model is given in Table II.\\
Here we note that the $T1(1,1)$ and $T2(1,1)$ models are the same, in 
particular, they have the same tree-level massless spectra and interactions.\\
(Note that $N1(1,1)$ and $N2(1,1)$ Narain models do not admit symmetric 
${\bf Z}_2$ orbifold.) \\
$\bullet$ The $T1(0,0)$ model. Start from the $N1(0,0)$ model and perform 
the following twists:
\begin{eqnarray}
 &&T_3 =(\theta,\theta,\theta\vert\vert 0, \theta, \theta)
 ({\cal P} \vert  2/3)~,\nonumber\\
 &&T_2=(0, \sigma,\sigma\vert\vert  e_1/2,\sigma, \sigma)
 (0^{15} \vert 0)~.\nonumber
\end{eqnarray}  
This model has $SU(2)_1 \otimes SO(10)_3 \otimes U(1)^4$ gauge symmetry. 
The massless spectrum of the $T1(0,0)$ model is given in Table III.\\
$\bullet$ The $T2(0,0)$ model. Start from the $N2(0,0)$ model and perform 
the same twists as in the $T1(0,0)$ model.
This model has $SU(2)_1 \otimes SO(10)_3 \otimes U(1)^3$ gauge symmetry. 
The massless spectrum of the $T2(0,0)$ model is given in Table III.\\
$\bullet$ The $T3(0,0)$ model. Start from the $N3(0,0)$ model and perform 
the same twists as in the $T1(0,0)$ model.
This model has $SU(2)_1 \otimes SO(10)_3 \otimes U(1)^2$ gauge symmetry. 
The massless spectrum of the $T3(0,0)$ model is given in Table III.\\
$\bullet$ The $T4(0,0)$ model. Start from the $N2(0,0)$ model and 
perform the following twists:
\begin{eqnarray}
 &&T_3 =(\theta,\theta,0\vert\vert 0, \theta, \theta)
 ({\cal P} \vert  2/3)~,\nonumber\\
 &&T_2=(0, \sigma,\sigma\vert\vert  e_1/2,\sigma, \sigma)
 (0^{15} \vert 0)~.\nonumber
\end{eqnarray}  
This model has $SU(2)_1 \otimes SO(10)_3 \otimes U(1)^3$ gauge symmetry. 
The massless spectrum of the $T4(0,0)$ model is given in Table IV.\\
$\bullet$ The $T5(0,0)$ model. Start from the $N3(0,0)$ model and perform 
the same twists as in the $T4(0,0)$ model. 
This model has $SU(2)_1 \otimes SO(10)_3 \otimes U(1)^2$ gauge symmetry. 
The massless spectrum of the $T5(0,0)$ model is given in Table IV.\\
$\bullet$ The $T6(0,0)$ model. Start from the $N4(0,0)$ model and perform 
the same twists as in the $T1(0,0)$ model.
This model has $SU(2)_1 \otimes SO(10)_3 \otimes U(1)^4$ gauge symmetry. 
The massless spectrum of the $T6(0,0)$ model is the same as that of the 
model $T1(0,0)$ with additional states coming from the $T_2$ sector. 
These additional states are shown in the first column of Table IX. \\
$\bullet$ The $T7(0,0)$ model. Start from the $N4(0,0)$ model, deform 
the corresponding Narain lattice in the direction $f=0$, $0<h<1$ 
(this breaks the gauge symmetry of the corresponding Narain model 
from $SU(3)^3 \otimes SO(10)^3 \otimes SO(2)$ down to
$SU(3)^2\otimes U(1)^2 \otimes SO(10)^3 \otimes SO(2)$), and perform the 
same twists as in the $T1(0,0)$ model.
This model has $SU(2)_1 \otimes SO(10)_3 \otimes U(1)^3$ gauge symmetry. 
The massless spectrum of the $T7(0,0)$ model is the same as that of the 
model $T2(0,0)$ with additional states coming from the $T_2$ sector. 
These additional states are shown in the second column of Table IX. \\
$\bullet$ The $T8(0,0)$ model. Start from the $N4(0,0)$ model, deform 
the corresponding Narain lattice (in the direction which breaks the 
gauge symmetry of the corresponding Narain model from 
$SU(3)^3 \otimes SO(10)^3 \otimes SO(2)$ down to
$SU(3)\otimes U(1)^4 \otimes SO(10)^3 \otimes SO(2)$), and 
perform the same twists as in the $T1(0,0)$ model. 
This model has $SU(2)_1 \otimes SO(10)_3 \otimes U(1)^2$ gauge symmetry. 
The massless spectrum of the $T8(0,0)$ model is the same as that of the 
model $T3(0,0)$ with additional states coming from the $T_2$ sector. 
These additional states are shown in the third column of Table IX. \\
$\bullet$ The $T9(0,0)$ model. Start from the $N4(0,0)$ model and perform 
the same twists as in the $T4(0,0)$ model.
This model has $SU(2)_1 \otimes SO(10)_3 \otimes U(1)^4$ gauge symmetry. 
The massless spectrum of the $T9(0,0)$ model is given in Table V.\\
$\bullet$ The $T10(0,0)$ model. Start from the $N4(0,0)$ model, deform 
the corresponding Narain lattice in the direction $f=0$, $0<h<1$ 
(this breaks the gauge symmetry of the corresponding Narain model from 
$SU(3)^3 \otimes SO(10)^3 \otimes SO(2)$ down to
$SU(3)^2\otimes U(1)^2 \otimes SO(10)^3 \otimes SO(2)$), and 
perform the same twists as in the $T4(0,0)$ model.
This model has $SU(2)_1 \otimes SO(10)_3 \otimes U(1)^3$ gauge symmetry. 
The massless spectrum of the $T10(0,0)$ model is given in Table V.\\
$\bullet$ The $T11(0,0)$ model. Start from the $N4(0,0)$ model, deform 
the corresponding Narain lattice (in the direction which breaks the 
gauge symmetry of the corresponding Narain model from 
$SU(3)^3 \otimes SO(10)^3 \otimes SO(2)$ down to
$SU(3)\otimes U(1)^4 \otimes SO(10)^3 \otimes SO(2)$), and 
perform the same twists as in the $T4(0,0)$ model.
This model has $SU(2)_1 \otimes SO(10)_3 \otimes U(1)^2$ gauge symmetry. 
The massless spectrum of the $T11(0,0)$ model is given in Table V.\\
(Here we can construct seven more three-family models via substituting 
the left-moving twist $(0,\theta,\theta)$ by $(e_1/3,0,\theta)$. 
This can be done in models $T1(0,0)$, $T6(0,0)$ and $T9(0,0)$; $T2(0,0)$, 
$T4(0,0)$, $T7(0,0)$ and $T10(0,0)$. In the first three cases the gauge 
group is $SU(2)_4 \otimes SO(10)_3 \otimes U(1)^4$. 
In the last four cases the gauge symmetry is 
$SU(2)_4 \otimes SO(10)_3 \otimes U(1)^3$. 
These seven models are not phenomenologically viable since they do not 
have non-abelian hidden sector. 
The $SU(2)_4 $ subgroup is a part of the horizontal symmetry 
in these models.)\\
$\bullet$ The $T1(1,0)$ model. Start from the $N1(1,0)$ model and 
perform the same twists as in the $T1(0,0)$ model.
This model has $SU(2)_1 \otimes SU(2)_3 \otimes SO(10)_3 \otimes U(1)^3$ 
gauge symmetry. 
The massless spectrum of the $T1(1,0)$ model is given in Table VI.\\
$\bullet$ The $T2(1,0)$ model. Start from the $N1(1,0)$ model and perform 
the following twists:
\begin{eqnarray}
 &&T_3 =(\theta,\theta,\theta\vert\vert e_1/3, 0, \theta)
 ({\cal P} \vert  2/3)~,\nonumber\\
 &&T_2=(0, \sigma,\sigma\vert\vert  e_1/2,\sigma, \sigma)
 (0^{15} \vert 0)~.\nonumber
\end{eqnarray}  
This model has $SU(2)_1 \otimes SU(2)_3 \otimes SO(10)_3 \otimes U(1)^3$ 
gauge symmetry. 
The massless spectrum of the $T2(1,0)$ model is given in Table VI.\\
$\bullet$ The $T3(1,0)$ model. Start from the $N1(1,0)$ model and perform 
the following twists:
\begin{eqnarray}
 &&T_3 =(\theta,0,\theta\vert\vert e_1/3, 0, \theta)
 ({\cal P} \vert  2/3)~,\nonumber\\
 &&T_2=(0, \sigma,\sigma\vert\vert  e_1/2,\sigma, \sigma)
 (0^{15} \vert 0)~.\nonumber
\end{eqnarray}  
This model has $SU(2)_1 \otimes SU(2)_3 \otimes SO(10)_3 \otimes U(1)^3$ 
gauge symmetry. 
The massless spectrum of the $T3(1,0)$ model is given in Table VII.\\
Here we note that the $T1(1,0)$ and $T3(1,0)$ models are the same, in 
particular, they have the same tree-level massless spectra and interactions.\\ 
$\bullet$ The $T4(1,0)$ model. Start from the $N1(1,0)$ model and perform 
the following twists:
\begin{eqnarray}
 &&T_3 =(\theta,0,\theta\vert\vert  0, \theta, \theta)
 ({\cal P} \vert  2/3)~,\nonumber\\
 &&T_2=(0, \sigma,\sigma\vert\vert  e_1/2,\sigma, \sigma)
 (0^{15} \vert 0)~.\nonumber
\end{eqnarray}  
This model has $SU(2)_1 \otimes SU(2)_3 \otimes SO(10)_3 \otimes U(1)^3$ gauge
symmetry. The massless spectrum of the $T4(1,0)$ model is given in Table VII.
Here we note that the $T2(1,0)$ and $T4(1,0)$ models are the same.

{}Here, a comment is in order.  We can choose asymmetric 
${\bf Z}_2$ twist since $SO(8)$ lattice 
admits an asymmetric ${\bf Z}_2$ orbifold. 
Thus, in the models with left-moving twists $(0,\theta,\theta)$ and
$(e_1/2,\sigma,\sigma)$ we can substitute them by $(0,e_1/3,\theta)$ 
and $(e_1/2,e_1/2, 0)$, respectively. 
This leads to the same models, {\em i.e.}, in this case symmetric 
and asymmetric orbifolds are equivalent. 
Similarly, in the models with left-moving twists $(e_1/3,0,\theta)$ and
$(e_1/2,\sigma,\sigma)$ we can substitute them by $(e_1/3,0,\theta)$ and 
$(e_1/2,e_1/2, 0)$, respectively. Again, this leads to the same models. \\
$\bullet$ The $T5(1,0)$ model. Start from the $N2(1,0)$ model and perform 
the following twists:
\begin{eqnarray}
 &&T_3 =(\theta,\theta,0\vert\vert e_1/3\vert \sqrt{2}/3,0,0,0)
 ({\cal P} \vert  2/3)~,\nonumber\\
 &&T_2=(0, \sigma,\sigma\vert\vert  e_1/2 \vert \sqrt{2}/2, 0,0,0)
 (0^{15} \vert 0)~.\nonumber
\end{eqnarray}  
In this model and the next, we use the 
$SO(8) \supset SU(2)^4$ basis for the left-moving $SO(8)$ momenta. 
Note that even multiples of $1/\sqrt{2}$ are the roots of $SU(2)$, 
whereas the odd multiples of $1/\sqrt{2}$ are the corresponding weights. 
The $SU(3)$ and $SO(8)$ left-moving momenta are separated by a single 
vertical line.
This model has $SU(2)^3_1  \otimes SO(10)_3 \otimes U(1)^4$ gauge symmetry. 
The massless spectrum of the $T5(1,0)$ model is given in Table VIII.\\
$\bullet$ The $T6(1,0)$ model. Start from the $N2(1,0)$ model and perform 
the following twists:
\begin{eqnarray}
 &&T_3 =(\theta,\theta,0\vert\vert 0 \vert 0,0, \sqrt{2}/3,\sqrt{2}/3)
 ({\cal P} \vert  2/3)~,\nonumber\\
 &&T_2=(0, \sigma,\sigma\vert\vert  e_1/2\vert \sqrt{2}/2, 0,0,0)
 (0^{15} \vert 0)~.\nonumber
\end{eqnarray}  
This model has $SU(2)^3_1  \otimes SO(10)_3 \otimes U(1)^4$ gauge symmetry. 
The massless spectrum of the $T6(1,0)$ model is given in Table VIII.\\
Here we note that the $T5(1,0)$ and $T6(1,0)$ models are the same. \\
Note that if we start from the $N1(1,0)$ model, then we cannot perform the 
same twists as in the models $T5(1,0)$ and $T6(1,0)$, since the $N1(1,0)$ 
Narain lattice does not admit the corresponding asymmetric 
${\bf Z}_3$ orbifold.\\
$\bullet$ The $T7(1,0)=T9(1,0)$, and $T8(1,0)=T10(1,0)$ models.  
Start from the $N2(1,0)$ model and perform the same twists as in 
models $T1(1,0)$, $T3(1,0)$, $T2(1,0)$ and $T4(1,0)$, respectively.
The massless spectra of the $T7(1,0)$, $T8(1,0)$, $T9(1,0)$ and $T10(1,0)$ 
models are the same as those of $T1(1,0)$, $T2(1,0)$, $T3(1,0)$ and 
$T4(1,0)$, respectively, with additional states coming from the $T_2$ sector. 
These additional states are the same for these four models, and are given 
in Table X.\\
(Here we note that the above four models $T7(1,0)$, $T8(1,0)$, $T9(1,0)$ 
and $T10(1,0)$ can also be constructed starting from the $N2(1,0)$ model 
and performing the ${\bf Z}_6$ orbifold where the ${\bf Z}_2$ orbifold 
is asymmetric. This gives the same set of models.
This is completely analogous to the case of the $T1(1,0)$, $T2(1,0)$, 
$T3(1,0)$ and $T4(1,0)$ models.)
{}To conclude, let us summarize the $3$-family GUT models: \\
    ~\\
$\bullet$ $T5(1,0)=T6(1,0)=SU(2)_1^3 \otimes SO(10)_3 \otimes U(1)^4$;\\
   ~\\
$\bullet$ $E1=E2=SU(2)_1 \otimes (E_6)_3 \otimes U(1)^3$;\\
   ~\\
$\bullet$ $T1(1,0)=T3(1,0)=SU(2)_1 \otimes SU(2)_3 \otimes SO(10)_3 
 \otimes U(1)^3$; \\
$\bullet$ $T2(1,0)=T4(1,0)=$ same as above; \\
$\bullet$ $T7(1,0)=T9(1,0)=$ same as above; \\
$\bullet$ $T8(1,0)=T10(1,0)=$ same as above; \\
   ~\\
$\bullet$ $T1(1,1)=T2(1,1)=SU(2)_1 \otimes SO(10)_3 \otimes U(1)^4$; \\
$\bullet$ $T1(0,0)=$ same as above; \\
$\bullet$ $T6(0,0)=$ same as above; \\
$\bullet$ $T9(0,0)=$ same as above; \\
   ~\\
$\bullet$ $T2(0,0)=SU(2)_1 \otimes SO(10)_3 \otimes U(1)^3$; \\
$\bullet$ $T4(0,0)=$ same as above; \\
$\bullet$ $T7(0,0)=$ same as above; \\
$\bullet$ $T10(0,0)=$ same as above; \\
   ~\\
$\bullet$ $T3(0,0)=SU(2)_1 \otimes SO(10)_3 \otimes U(1)^2$; \\
$\bullet$ $T5(0,0)=$ same as above; \\
$\bullet$ $T8(0,0)=$ same as above; \\
$\bullet$ $T11(0,0)=$ same as above; \\
       ~\\
{}It is clear that the last two sets of models can be obtained from either 
of the former two sets of models via spontaneous symmetry breaking. 
We shall discuss this in the next section.

{}Finally, let us summarize the new versus old results presented in this paper.
The models $T1(1,0) $ and $ T1(0,0)$ were first presented in Ref \cite{three}. 
In Ref \cite{kt}, the constructions of these models were discussed in detail. 
Ref \cite{kt} also presents the models $E1$, $T1(1,1)$,
$T2(0,0)$, $T3(0,0)$ and $T2(1,0)$. 
The rest of the models classified in this paper are new. 
This concludes the classification of $3$-family $SO(10)$ and $E_6$ string
models.

\section{Moduli Space of $SO(10)$ and $E_6$ Models}

\bigskip

{}Here, we point out that all but the unique $SO(10)$ model classified 
in this paper are connected by (classically) flat moduli. In section III,
the moduli space of the $N=4$ supersymmetric models is discussed. That 
moduli space provides an underlying structure to the moduli space for
the orbifold models constructed. 
In subsection A here, we give a discussion of these flat directions in 
the language of effective field theory.  In particular, we identify 
the massless scalar fields in these models whose vevs correspond 
to these flat moduli. In subsection B, we give a stringy description of this
moduli space. Fig. 1. summarizes the discussion in this section.

\subsection{Massless Scalars and Flat Directions}

{}Let us study the effective field theory limit of the string models 
we have constructed. There are scalar fields in the massless spectra of 
these models, and some of them correspond to flat directions.  
In our following discussion we implicitly use our convention, 
which was mentioned in the Introduction, for distinguishing different models. 
The two models are considered to be the same if their tree-level massless 
spectra and/or interactions are the same. 
Thus, when  connecting different models with flat moduli we do not have 
to worry about the precise matching of the corresponding heavy string states 
(generically they will not match, but there will be a point in the 
moduli space where the matching will be exact; then the corresponding 
models are connected via this point, and the massless spectra for 
continuously many models could be the same while the massive spectra are not; 
an example of this will be given in the next subsection). 

{}Thus, consider Fig. 1. There we have already identified the pairs of
models that are the same ({\em e.g.}, $T1(1,1)=T2(1,1)$;
see section IV for details).
The symbols $a$, $b$, $c$, $d$ and $e$ stand for the fields whose vevs are
the corresponding flat directions. Thus, $c$ is the triplet Higgs ({\em i.e.},
(${\bf 1},{\bf 3}, {\bf 1})(0,0,0)_L$) of the $SU(2)_3$ subgroup in
models $T1(1,0) =T3(1,0)$, $T2(1,0)=T4(1,0)$, $T7(1,0)=T9(1,0)$ and
$T8(1,0)=T10(1,0)$.
Upon giving this field a vev in the direction corresponding to breaking
$SU(2)_3 \supset U(1)$, all four models have the massless spectrum of
the model $T1(1,1)=T2(1,1)$. Similarly, $E1=E2$ model goes into
$T1(1,1)=T2(1,1)$ model upon the adjoint Higgs $e$ of $(E_6)_3$
({\em i.e.}, $({\bf 1}, {\bf 78}) (0,0,0)_L$) acquiring a vev in the
direction such that $(E_6)_3 \supset SO(10)_3 \otimes U(1)$
breaking occurs. Note that in a similar fashion we could connect
some models with, say, $SU(5)$ gauge symmetry to the models
classified in this paper.

{}Next, we briefly discuss the remaining flat directions shown in Fig. 1.
The field $b$ is a doublet of $SU(2)_3$ charged under the corresponding
$U(1)$ in models $T1(1,0) =T3(1,0)$, $T2(1,0)=T4(1,0)$, $T7(1,0)=T9(1,0)$
and $T8(1,0)=T10(1,0)$ ({\em e.g.}, $({\bf 1}, {\bf 2},
{\bf 1})(0,-3,0)_L$ in the $T1(1,0)$ model).
Upon giving this field a vev, the gauge subgroup $SU(2) \otimes U(1)$
breaks down to $U(1)$ (the latter is a mixture of the original $SU(2)$
and $U(1)$), and the corresponding models can be read off from Fig. 1.
If a field $d$ gets a vev in models $T1(1,0) =T3(1,0)$, $T2(1,0)=T4(1,0)$,
$T7(1,0)=T9(1,0)$ and $T8(1,0)=T10(1,0)$
(in these models $d$ is  $({\bf 2},{\bf 2}, {\bf 1})(0,0,0)_L$),
the $SU(2)_1 \otimes SU(2)_3$ gauge symmetry is broken down to $SU(2)_4$,
and the corresponding models do not have a hidden sector.
Finally, the fields $a$ in the $T(0,0)$ models ({\em e.g.}, $({\bf 1},
{\bf 1})(0,-6,0,0)_L$ in the $T1(0,0)$ model) are singlets charged under
corresponding $U(1)$s,
so that upon acquiring a vev, the $U(1)$ gauge symmetry is broken.

{}The above description of these models in terms of the moduli space is
useful in the sense that certain conclusions can be made for all of these
models at once. Thus, let us for a moment ignore the string threshold
corrections and possible stringy non-perturbative effects.
Then we can analyze these models in the field theory context.
The dynamics of the hidden sector then is determined by the number of
doublets of the $SU(2)_1$ hidden sector.
The hidden sector is asymptotically-free only for the points in the
moduli space where the number of doublets is less than 12
(this would correspond to 6 "flavors"). There are only 9 such models, namely,
$E1$, $T1(1,1)$, $T1(1,0)$, $T2(1,0)$, and $T1-5(0,0)$.
The first two models, namely, $E1$ and $T1(1,1)$ have 2 doublets,
which corresponds to 1 "flavor".
In the case of models $T1(1,0)$, $T2(1,0)$, and $T1-5(0,0)$ we have 6
doublets, corresponding to 3 "flavors".

{}Finally, we comment on the $T5(1,0)=T6(1,0)$ model.
This model does not seem to be connected to the other 17 models considered
in this paper in any simple way. This can be readily seen from the fact
that the structure of generations in the $T5(1,0)=T6(1,0)$ model is
different from that of the rest of the models.
Such a connection would involve breaking the hidden sector,
so the corresponding models (via which the connection is possible)
would not be in our classification.

\subsection{Stringy Flat Directions}

{}The above discussion is carried out in the language of the low energy 
effective field
theory. If there are flat directions that can be described as vevs
of the corresponding scalars, there must be a string theory description
for them as well. In practice, however, finding such a description
explicitly may be involved, especially in complicated models such
as those classified in this paper.
Here we give an example of such a description in the case of our models.

{}Let us start from the Narain model $N(h,0)$, $0\leq h \leq 1$
(see section III for details) and turn on the following Wilson lines:
\begin{eqnarray}
 &&U_1 =(e_1/2,a_1,b_1 \vert\vert 0,c_1,d_1)({\bf s}\vert {\bf 0}\vert 
 {\bf 0} \vert {\overline S})~,\nonumber \\
 &&U_2 =(e_2/2,a_2,b_2 \vert\vert 0,c_2,d_2)({\bf 0}\vert {\bf s}\vert 
 {\bf 0} \vert {\overline S})~.\nonumber
\end{eqnarray}
Here
\begin{eqnarray}
  (a_1, b_1) =&&(-h{\tilde E}^6 +E_3 +E_5 )/2~,~~~0\leq h\leq 1/2 ~,\nonumber\\
  (a_1, b_1) =&&((h-1){\tilde E}^6 +E_3 +E_5 )/2~,~~~1/2\leq h\leq 1 ~,
 \nonumber \\
  (a_2, b_2) =&&( h{\tilde E}^5 +E_4 +E_6 )/2~,~~~0\leq h\leq 1/2 ~,\nonumber\\
  (a_2, b_2) =&&(- (h-1){\tilde E}^5 +E_4 +E_6 )/2~,~~~1/2\leq h\leq 1 ~, 
 \nonumber\\
  (c_1, d_1) = &&-h{\tilde E}^6/2~,~~~0\leq h\leq 1/2 ~,\nonumber \\
  (c_1, d_1) = &&(h-1){\tilde E}^6/2~,~~~1/2\leq h\leq 1 ~,\nonumber \\
  (c_2, d_2) = &&h{\tilde E}^5/2~,~~~0\leq h\leq 1/2 ~,\nonumber \\
  (c_2, d_2) = &&-(h-1){\tilde E}^5/2~,~~~1/2\leq h\leq 1 ~,\nonumber
\end{eqnarray}

{}Let ${\tilde N}(h,1)$ be the corresponding Narain model.
The gauge symmetry of this model is
$SU(3)\otimes {\tilde R}(h,0) \otimes SO(10)^3 \otimes SO(2)$.
Here ${\tilde R}(0,0)=SU(3)\otimes SU(3)$,
${\tilde R}(0<h<1,0)=SU(3)\otimes U(1)^2$, and ${\tilde R}(1,0)=SO(8)$.
Now let us start from the ${\tilde N}(h,1)$ model and perform the same twists
as in the $T1(0,0)$ model (see section IV for details).
The corresponding model depends on the value of the modulus $h$.
For $h=0$ we have the $T1(0,0)$ model. For $h=1$ we have the $T1(1,0)$ model.
For $0<h<1$ the resulting model has the massless spectrum which is the same
as that of the $T2(0,0)$ model. Thus these three models are indeed
connected by a continuous parameter $h$.
In the previous subsection we described this modulus as vev of the
corresponding scalar fields in the effective field theory language.

\newpage
\begin{figure}[t]
\epsfxsize=16 cm
\epsfbox{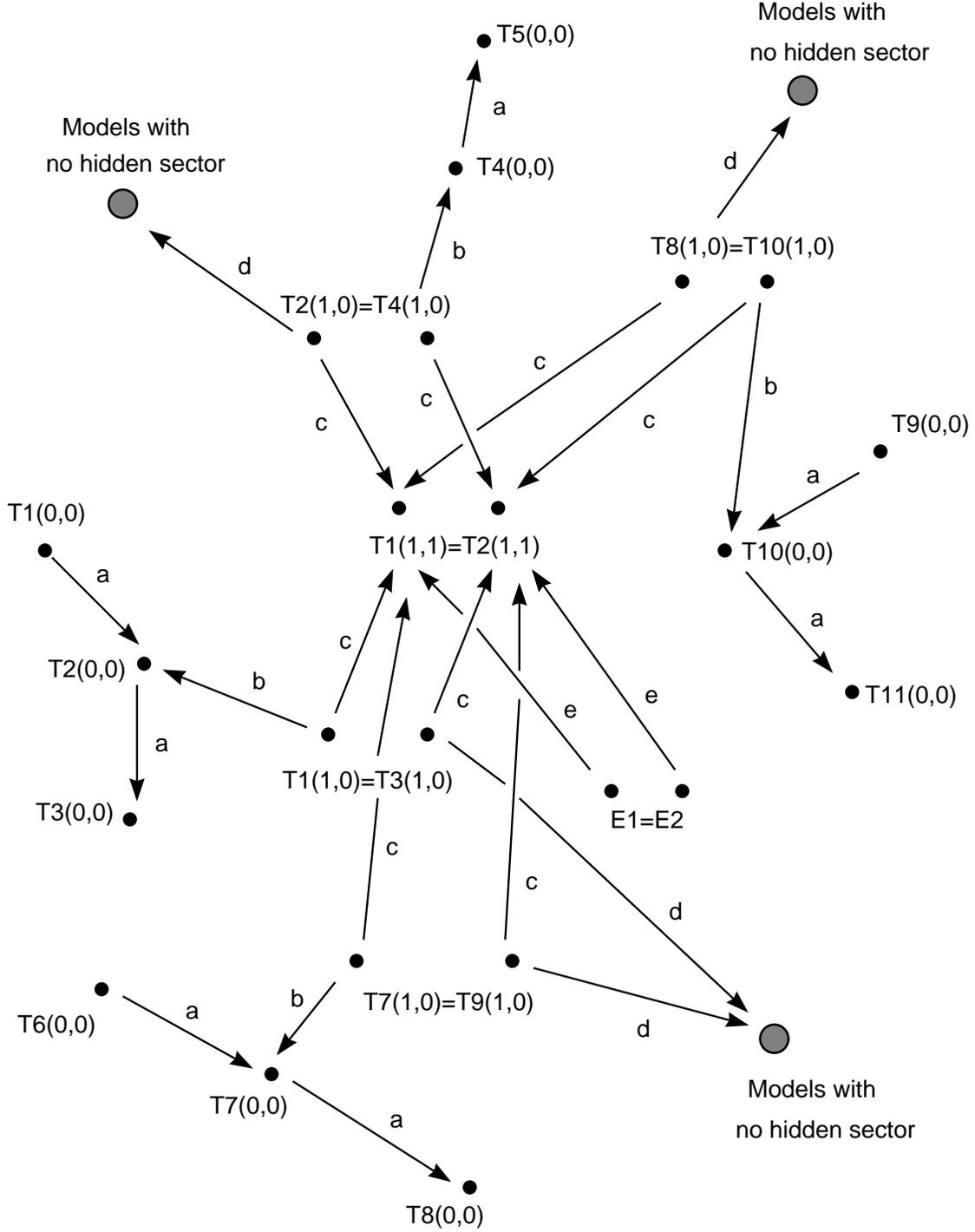}
\caption{The space of models. All models are connected by flat directions. 
Here $a$, $b$, $c$, $d$ and $e$ stand for the fields whose vevs are the 
corresponding flat directions. See section V for details.}
\end{figure}

\newpage

\section{Conclusions}

{}In this paper we have given a classification of $3$-family $SO(10)$ 
and $E_6$ grand unified models in string theory within the framework of 
conformal field theory and asymmetric orbifolds.
Is it possible that our classification does not exhaust all 
such models that can be constructed within conformal field theory? 
This could happen if some $(2,0)$ Calabi-Yau compactifications can 
give $3$-family $SO(10)$ and $E_6$ models
and yet not be connected to our models (which in turn are certain singular 
points in the moduli space of $(2,0)$ Calabi-Yau compactifications) 
by a modulus
({\em e.g.}, this could happen if, in order to connect these models, one 
has to pass through certain singularities). However, 
level-3 $SO(10)$ and $E_6$ Kac-Moody algebras only have diagonal embedding, 
corresponding to precisely modding out by a ${\bf Z}_3$ outer-automorphism. 
So, this implies that our list exhausts all the possible models of this 
type that admit free field realization. 
We also pointed out that, say, level-4 
$SO(10)$ models are very likely to have even numbers of chiral families, 
and that there is no room for the hidden sector in such models. 
This is true for the embedding
$SO(10)_4 \subset SO(10)^4$, although to the best of our knowledge there 
have been no previous attempts to construct such models in the literature. 
It would be interesting to explore this particular avenue since 
higher-level $SO(10)$ Kac-Moody algebras
(as well as $E_6$ Kac-Moody algebras at levels higher than 3) do not 
appear to have embeddings in the corresponding level-1 algebras with 
central charges less than or equal 22. 
There is another possibility: $SO(10)_4 \subset SU(16)$. 
Unfortunately, this embedding does not seem to have a simple orbifold 
realization, at least not a ${\bf Z}_3$ orbifold realization, so 
it remains an open question if such models can be constructed.  
Even if such a model can be constructed, it will not have the ${\bf Z}_3$
structure needed for three families.

\acknowledgments

\bigskip

{}We would like to thank Michael Bershadsky, Albion Lawrence, Pran Nath, 
Lisa Randall, Gary Shiu, Tom Taylor and Yan Vtorov-Karevsky for discussions.
The research of S.-H.H.T. was partially supported by National Science 
Foundation.
The work of Z.K. was supported in part by the grant NSF PHY-96-02074, and 
the DOE 1994 OJI award. Z.K. would also like to thank Mr. Albert Yu 
and Mrs. Ribena Yu for financial support.

\newpage

\appendix \section{Model-building}
\bigskip

{}In this appendix, let us give a brief outline/introduction to some of the key
elements in the construction of asymmetric orbifold models, and a few 
illustrations on how to obtain the massless spectra. Starting from an 
$N=4$ Narain model, a new model can be generated by performing a consistent
set of twists and shifts (of the momentum lattice) on it.
These twists and shifts must be consistent with the lattice symmetry.
The rules of consistent asymmetric orbifold model construction can then
be written as constraints on them, as done in Ref\cite{kt}.
These constraints follow from the one-loop modular invariance,
world-sheet supersymmetry, and the physically sensible projection conditions.
All sets of shifts and twists given in the text,
${\em i.e.}$, the Wilson lines $(U_1, U_2)$, and the ${\bf Z}_6$ orbifolds
$(T_3, T_2)$, have been checked to satisfy these consistency constraints.

{}In principle, the spectrum of the resulting model can be
obtained from the spectrum generating formula\cite{kt}. 
In some cases, parts of the massless spectrum can be obtained by imposing 
only the level-matching condition and the anomaly-free condition. 
Although this approach is simple, it does not yield
all the discrete quantum numbers of the particles, which are important for
couplings. For these properties, one should go back to the spectrum
generating formula. 
Sometimes, the identification of the quantum numbers
of the particles can be quite non-trivial. Fortunately, whenever this happens, 
we are able to find an alternative construction of the same model where 
those particles are clearly identified. This provides powerful checks.

{}The constructions of all the models are very similar. To be explicit, let 
us illustrate the steps with the construction of the $T1(1,0)$ model.
We shall start with the N=4 supersymmetric Narain model with the
$SU(3) \otimes SO(8) \otimes SO(32)$ gauge symmetry. Here,
the gauge symmetry $SU(3) \otimes SO(8)$ comes from the $6$ compactified
dimensions. This  model corresponds
to the point $(h,f)=(1,0)$ in the moduli space discussed in section III,
${\em i.e.}$, $R(1,0)=R(0,1)=SU(3) \otimes SO(8)$.
The following Wilson lines are introduced to act on this model,
\begin{eqnarray}
 &&U_1 =(e_1/2,a_1,b_1\vert\vert 0,0,0)({\bf s}\vert {\bf 0}\vert {\bf 0}\vert
 {\overline    S})~,\nonumber\\
 &&U_2 =(e_2/2,a_2,b_2 \vert\vert  0,0,0)({\bf 0}\vert {\bf s}\vert
 {\bf 0}\vert {\overline S})~.
\end{eqnarray}
The first three entries correspond to the right-moving complex world-sheet
bosons. The next three entries correspond to the left-moving complex
world-sheet bosons. Together they form the six-torus.
Here the four component vectors $(a_1,b_1)$ and $(a_2,b_2)$ are the spinor
${\bf  s}$ and conjugate $ {\bf  c}$ weights of $SO(8)$, respectively.
The remaining 16 left-moving world-sheet bosons generate the
${\mbox{Spin}}(32)/{\bf Z}_2$ lattice.
The $SO(32)$ shifts are given in the $SO(10)^3 \otimes SO(2)$ basis.
In this basis, ${\bf 0}$($0$) stands for the null vector,
${\bf v}$($V$) is the vector weight, whereas
${\bf s}$($S$) and ${\overline {\bf s}}$(${\overline S}$) are the
spinor and anti-spinor weights of $SO(10)$($SO(2)$). (For $SO(2)$, $V=1$,
$S=1/2$ and ${\overline S}=-1/2$.) 
Under $SO(32)\supset SO(10)^3 \otimes SO(2)$, we have
\begin{eqnarray}\label{aa1}
 {\bf 0}&=&({\bf 0},{\bf 0},{\bf 0},0)+({\bf v},{\bf v},{\bf v},V)+
 ({\bf 0},{\bf 0},{\bf v},V)+({\bf v},{\bf v},{\bf 0},0)+ \nonumber \\
 &&({\bf 0},{\bf v},{\bf 0},V)+({\bf v},{\bf 0},{\bf v},0)+
 ({\bf v},{\bf 0},{\bf 0},V)+({\bf 0},{\bf v},{\bf v},0). 
\end{eqnarray}
Since this $(U_1, U_2)$ set involves shifts only (${\em i.e.}$, no twist),
the resulting model still has $N=4$ supersymmetry. The states in the
original model that will remain in the new model are those
that are invariant under these shifts. Of the original ones,
the $SU(3)\otimes SO(8)$ gauge bosons remain untouched,
since there is no shift on their momentum lattices. To see what happens to 
the $SO(32)$ gauge symmetry,
it is convenient to consider $SO(32)$ weights in terms of
$SO(10)^3 \otimes SO(2)$ weights.
The $SO(10)^3 \otimes SO(2)$ gauge bosons come from the
$({\bf 0},{\bf 0},{\bf 0},0)$ weight lattice in Eq. (\ref{aa1}) and the 
Cartan generators, while the last $7$ terms in Eq. (\ref{aa1}) yields the 
remaining gauge bosons of $SO(32)$.
Since massless states coming from these last $7$ terms are not invariant
under the shift vectors $(U_1, U_2)$, they are projected out.
So the resulting gauge symmetry of this new model is 
$SU(3)\otimes SO(8)\otimes SO(10)^3 \otimes SO(2)$.
This truncated model is no longer a consistent string model. To restore
string consistency, the vectors $(U_1, U_2)$ introduce new sectors 
(namely, the $U_1$, the $U_2$ and the $U_1 + U_2$ sectors) of
particles (but they do not introduce new gauge bosons). Together, 
they yield the new Narain model, to be called the $N1(1,0)$ model.

{}Next we can perform a ${\bf Z}_6$ orbifold on the above $N1(1,0)$ model.
It is easier to follow the construction by decomposing the ${\bf Z}_6$
orbifold into a ${\bf Z}_3$ twist and a ${\bf Z}_2$ twist:
\begin{eqnarray}
 &&T_3 =(\theta,\theta,\theta\vert\vert 0, \theta, \theta)
 ({\cal P} \vert  2/3)~,\nonumber\\
 &&T_2=(0, \sigma,\sigma\vert\vert  e_1/2,\sigma, \sigma)
 (0^{15} \vert 0)~.
\end{eqnarray}
Each $\theta$ in $T_3$ is a ${\bf Z}_3$ twist (that is, a $2\pi /3$ rotation)
that acts only on a complex world-sheet boson. So the first $\theta$ acts on
the right-moving part of the $\Gamma^{2,2}$ (${\em i.e.}$, the
$SU(3)$) lattice and the corresponding oscillator excitations, while
the left-moving part is untouched. This is an asymmetric orbifold.
The $\Gamma^{4,4}$ (${\em i.e.}$, the $SO(8)$) lattice is twisted
symmetrically by the ${\bf Z}_3 \otimes {\bf Z}_3$ $\theta$ twist.
The three $SO(10)$s are permuted by the action of the
${\bf Z}_3$ outer-automorphism twist ${\cal P}$:
$\phi^I_1 \rightarrow
\phi^I_2 \rightarrow \phi^I_3 \rightarrow \phi^I_1$, where the real bosons
$\phi^I_p$, $I=1,...,5$, correspond to the $p^{\mbox{th}}$ $SO(10)$
subgroup, $p=1,2,3$. We can define new bosons $\varphi^I \equiv {1\over
\sqrt{3}}(\phi^I_1 +\phi^I_2 +\phi^I_3)$; the other ten real bosons are
complexified via linear combinations $\Phi^I \equiv {1\over
\sqrt{3}}(\phi^I_1 +\omega\phi^I_2 +\omega^2 \phi^I_3)$ and
$(\Phi^I)^\dagger \equiv {1\over \sqrt{3}}(\phi^I_1 +\omega^2\phi^I_2
+\omega \phi^I_3)$, where $\omega =\exp(2\pi i /3)$.
Under ${\cal P}$, $\varphi^I$ is invariant,
while $\Phi^I$ ($(\Phi^I)^\dagger$) are eigenstates with
eigenvalue $\omega^2$ ($\omega$). Finally, string consistency requires the 
inclusion of the $2/3$ shift in the $SO(2)$ lattice. This simply changes the 
radius of this left-moving world-sheet boson.

{}Notice that all states in the untwisted sector of the $T1(1,0)$ model must 
be invariant under the $T_3$ and $T_2$ twists. However, they need not be 
invariant under the ${\bf Z}_3$ outer-automorphism alone. The only states that must 
be invariant under the ${\bf Z}_3$ outer-automorphism are the states that are 
already invariant under other parts of the $T_3$ twist.
These include the grand unified gauge bosons. 
So, out of the $SO(10)^3$ gauge bosons, only the states that are invariant 
under the ${\bf Z}_3$ outer-automorphism are kept. They yield the $SO(10)_3$.

{}The model that results from twisting by
the above $T_3$ twist has $N=1$ space-time supersymmetry.
This can be easily seen by counting the number of gravitinos. This counting
depends only on the right-movers, which
is identical to the original symmetric orbifold model \cite{orb}.
All the gauge bosons come from the untwisted sector, and the gauge group
is $SU(3)_1 \otimes SU(3)_3 \otimes SO(10)_3 \otimes U(1)$.
The twisted sectors give rise to chiral matter fields of $SO(10)_3$.
The asymmetric ${\bf Z}_3$ twist $(\theta \vert\vert 0)$
in $\Gamma^{2,2}$ contributes only a factor of $1$ to the number of
fixed points as the factor $3$ contributed by its right mover is
cancelled against the volume factor of the corresponding invariant 
sublattice, which is $\Gamma^2$. Similarly, the ${\bf Z}_3$ 
outer-automorphism twist contributes only one fixed point. 
This follows from the form of the invariant sublattice.
The only non-trivial contribution to the number of fixed points in the 
twisted sectors comes from the symmetric ${\bf Z}_3$ twist in $\Gamma^{4,4}$.
This twist contributes $9=3_R \times 3_L$ fixed points.

{}To see how level-matching works, let us illustrate with the scalar 
superpartners of these matter fields coming from the ${\bf Z}_3$ twisted 
sector, namely, the $T3$ sector. The right-moving ground state energy of
this sector is given by
\begin{equation}
        E_R = 3({1\over 2}(1/3)(1 - 1/3) + {1\over 2}(1/3)^2) - {1/2} = 0
\end{equation}
where the first term gives the conformal dimension of the ${\bf Z}_3$
fixed points, and the second term gives the conformal dimension of the
corresponding world-sheet fermions, whose spin-structures are dictated
by world-sheet supersymmetry. The factor $3$ comes from the
$3$ complex world-sheet bosons and fermions.

{}The left-moving ground state energy of this sector can be calculated 
in a number of ways. 
In the above basis of $\Phi^I$ and $\varphi^I$, $I=1,...,5$,
the ${\bf Z}_3$ outer-automorphism twist ${\cal P}$ can be written as
\begin{equation}
        {\cal P}=(\theta ^5, 0^5)
\end{equation}
That is, ${\cal P}$ is equivalent to a ${\bf Z}_3$ twist on the $5$
complex world-sheet bosons $\Phi^I$, while leaving the $\varphi^I$ untouched. 
So the left-moving ground state energy is given by
\begin{equation}
        E_L = (2+5){1\over 2}(1/3)(1 - 1/3) - 1  = -2/9.
\end{equation}
To identify the $SO(10)_3$ chiral families, we are interested in finding
the momenta ${\bf p}$ inside the dual of the ${\bf Z}_3$-invariant sublattice
$\Gamma^6$ made of $\varphi^I$ and $SO(2)$, ${\em i.e.}$, 
${\bf p}=({\bf q}/ \sqrt{3} \vert Q + {2\over 3})$. That is, we want
to find ${\bf p}$ such that 
\begin{equation}
        E_L = -{2\over 9} +{1\over 6}{\bf q}^2 +{1\over 2}(Q+{2\over 3})^2 = 0.
\end{equation}
This is satisfied for $({\bf q} \vert Q)$ = $({\bf s} \vert -1/2)$,
$({\bf v} \vert -1)$ and $({\bf 0} \vert 0)$, since the conformal 
dimension of ${\bf v}$ is $1/2$ and that of ${\bf s}$ is $5/8$.
We see that level-matching requires the $2/3$ shift in the $SO(2)$ lattice. 

{}Recall that there are $9$ fixed points in this ${\bf Z}_3$ twisted sector.
The left-moving fixed points fall under irreps of the $SU(3)_3$; so
we have three copies (due to the three right-moving fixed points) of
massless states in the $SU(3)_1\otimes SU(3)_3 \otimes SO(10)_3 \otimes U(1)$ 
irreps $({\bf 1}, {\overline {\bf 3}}, {\bf 16})(-1)_L$,
$({\bf 1}, {\overline {\bf 3}},
{\bf 10})(+2)_L$, $({\bf 1}, {\overline {\bf 3}}, {\bf 1})(-4)_L$
(Here we give the $U(1)$ charge in parentheses, and its normalization
is $1/6$). Note that the $SU(3)_3$ anomaly coming from these states is precisely cancelled by
the $3$ copies of $({\bf 1}, {\bf 10}, {\bf 1})(0)_L$ coming 
from the untwisted sector.

{}To cut the number of families to $3$, the $T_2$ twist is introduced.
The $\sigma$ in the $T_2$ twist is a $\pi$ rotation of the
corresponding two chiral world-sheet bosons. Thus, $\sigma$ is a
${\bf Z}_2$ twist.
The left-moving momenta of $\Gamma^{2,2}$ are shifted by
$e_1/2$ (${\em i.e.}$, half a root vector), while the $\Gamma^{16}$
is left untouched. So this model has 
$SU(2)_1 \otimes SU(2)_3 \otimes SO(10)_3 \otimes U(1)^3$ gauge symmetry.

{}Of the $9$ fixed points in the ${\bf Z}_3$ twisted sector, 
the one at the origin is invariant under this
${\bf Z}_2$ twist. The remaining $8$ fixed points form $4$ pairs, and
the ${\bf Z}_2$ twist permutes the $2$ fixed points in each pair.
Forming $4$ symmetric and $4$ antisymmetric
combinations, we have $9=5(1)+4(-1)$ (where the
${\bf Z}_2$ phases are given in parentheses); that is, $5$ of the original
$9$ are invariant under the ${\bf Z}_2$ twist.
Since there is no relative
phase between the $T2$ and $T3$ sectors, these $5$
copies of the $SO(10)_3$ chiral matter fields survive, while the other
$4$ are projected out. The $SU(2)_3$ quantum numbers of these $5$ left-handed 
families can be easily worked out. They, and the remaining
massless spectrum of the resulting orbifold, the $T1(1,0)$ model, 
is given in Table VI.

\section{Non-Chiral Cases}
\bigskip

{}In this section we briefly discuss models that are not chiral. In particular, we explain why
it is not possible to start from an $N=1$ non-chiral model which is obtained with a single ${\bf Z}_3$ twist, and introduce chiral fermions by ${\bf Z}_2$ orbifolding it. First, consider the following example. Start from the $N1(0,0)$ model and perform the following ${\bf Z}_3$ twist:
\begin{eqnarray}
 T_3 =(\theta,\theta,\theta\vert\vert 0, 0, \theta)
 ({\cal P} \vert  0)~,\nonumber
\end{eqnarray}  
This model has $N=1$ space-time supersymmetry and $SU(3)_1\otimes SU(3)_1
 \otimes U(1)^2 \otimes SO(10)_3 \otimes SO(2)$ gauge symmetry. Note that in the above twist we have not shifted the $SO(2)$ lattice. This shift was the source of chirality of fermions coming from the twisted sectors of all the models considered in the main text, {\em i.e.}, it correlates the space-time helicity of a state in the twisted sector with the gauge quantum numbers. Thus, in the above model, the fermions from the twisted sectors are not chiral. In particular, the $T_3$ sector gives rise to $3$ left-handed families of  fermions in ${\bf 16}$ of $SO(10)_3$, and also $3$ right-handed families of  fermions in ${\bf 16}$ of $SO(10)_3$. These combine into $3$ non-chiral families of fermions transforming in the spinor representation of $SO(10)_3$. Suppose now we would like to project the right-handed fermions out by adding a ${\bf Z}_2$ twits. To do this, we will have to correlate the $SO(10)_3\otimes SO(2)$ quantum numbers with the right-moving word-sheet fermion quantum numbers in this ${\bf Z}_2$ twist
. This, however, as we already explained in the beginning of section IV, would destroy the ${\bf Z}_3$ symmetry of the lattice, so that the resulting model would not be a consistent orbifold model.

{}Note that our choice of ${\bf Z}_3$ twists in the main text was dictated by the requirements of $N=1$ space-time supersymmetry, presence of  chiral fermions, and the level-matching requirement. These twists then fix the form of the additional ${\bf Z}_2$ twists that must be added to obtain models with three families. Here we note that no other order twists (say, ${\bf Z}_4$) would be compatible with the ${\bf Z}_3$ twists. 

{}Finally, we point out one interesting feature of the non-chiral model given in this appendix.
This model has three families of adjoints of $SO(10)_3$ coming from the untwisted sector. These are neutral under all the other gauge groups. There are, however, three additional adjoints coming from the twisted sector. These are charged under the $U(1)^2$ (but they are neutral under $SU(3)_1\otimes SU(3)_1\otimes SO(2)$). This is an example of a model, although this is only a toy model, where higher dimensional Higgs fields can carry additional gauge quantum numbers. Such fields can only come from the twisted sector, whereas, say, adjoints that come from the untwisted sector are always neutral under all the other gauge groups \cite{dien}. 


\begin{table}[t]
\begin{tabular}{|c|l|l|}
 &$E1$ &  $E2$\\
 M & $SU(2) \otimes E_6 \otimes U(1)^3$ & $SU(2) \otimes E_6 \otimes U(1)^3$ 
      \\ \hline
 & & \\
   & $ ({\bf 1},{\bf 78})(0,0,0)_L$ & $ ({\bf 1},{\bf 78})(0,0,0)_L$ \\
  $U$ & $2 ({\bf 1},{\bf 1})(0,-{3},\pm 3)_L$ & $ ({\bf 1},{\bf 1})(\pm 3,+{3},0)_L$ \\
   & $ ({\bf 1},{\bf 1})(0,+6,0)_L$ & $ ({\bf 1},{\bf 1})(0,+6,0)_L$ \\
    & & $({\bf 2}, {\bf 1})(0,0,\pm 3)_L$ \\
  & & \\
  \hline
 & &  \\
   &  $ ({\bf 1},{\bf 27})(0,-{2},0)_L$ &  $ ({\bf 1},{\bf 27})(0,-{2},0)_L$ \\
    $T3$  & $2({\bf  1}, {\bf 27})(0,+1,\pm1)_L$ & $ ({\bf 1},{\overline {\bf 27}}) (\pm 1,-1,0)_L$ \\
 & & \\
 \hline
  & & \\
  T6  &  $ ({\bf 1},{\overline {\bf 27}}) (\pm 1,-1,0)_L$ & $2({\bf  1}, {\bf 27})(0,+1,\pm1)_L$  \\
    & &  \\
 \hline
 & &  \\
   $T2$ & $({\bf 2},{\bf 1})(0,0,{\pm 3})_L$ & $2 ({\bf 1},{\bf 1})(0,-{3},\pm 3)_L$ \\
    &  $ ({\bf 1},{\bf 1})(\pm {3},+{3},0)_L$ & \\
  & &\\
   \hline
 & &  \\
 $U(1)$ & $(1/ \sqrt{6}, ~1/{3\sqrt{2}}, ~1/\sqrt{6})$ & $(1/ \sqrt{6}, ~1/{3\sqrt{2}},  ~1/\sqrt{6})$\\
\end{tabular}
\caption{The massless spectra of the two $E_6$  models $E1$ and $E2$ both with gauge symmetry $SU(2)_1 \otimes (E_6)_3\otimes U(1)^3$.
The $U(1)$ normalization
radii are given at the bottom of the Table.
The gravity, dilaton and gauge supermultiplets are not shown.}

\end{table}

\begin{table}[t]
\begin{tabular}{|c|l|l|}
 &$T1(1,1)$ &  $T2(1,1)$\\
 M & $SU(2) \otimes SO(10)\otimes U(1)^4$ & $SU(2) \otimes SO(10) \otimes U(1)^4$ 
      \\ \hline
 & & \\
   & $ ({\bf 1},{\bf 45})(0,0,0,0)_L$ & $ ({\bf 1},{\bf  45})(0,0,0,0)_L$ \\
   & $ ({\bf 1},{\bf 1})(0,0,0,0)_L$ & $ ({\bf 1},{\bf  1})(0,0,0,0)_L$ \\ 
  $U$ & $2 ({\bf 1},{\bf 1})(0,-{3},\pm 3,0)_L$ & $ ({\bf 1},{\bf 1})(\pm 3,+{3},0,0)_L$ \\
   & $ ({\bf 1},{\bf 1})(0,+6,0,0)_L$ & $ ({\bf 1},{\bf 1})(0,+6,0,0)_L$ \\
    & & $({\bf 2}, {\bf 1})(0,0,\pm 3,0)_L$ \\
  & & \\
  \hline
 & &  \\
   &  $ ({\bf 1},{\bf  16})(0,-{2},0,-1)_L$ &  $ ({\bf 1},{\bf  16})(0,-{2},0, -1)_L$ \\
   &   $ ({\bf 1},{\bf  10})(0,-{2},0,+2)_L$ &  $ ({\bf 1},{\bf  10})(0,-{2},0,+2)_L$ \\
   &   $ ({\bf 1},{\bf  1})(0,-{2},0,-4)_L$ &  $ ({\bf 1},{\bf 1})(0,-{2},0,-4)_L$ \\
 $T3$  
   & $2({\bf  1}, {\bf  16})(0,+1,\pm1,-1)_L$ &  $ ({\bf 1},{\overline {\bf 16}}) (\pm 1,-1,0,+1)_L$ \\
   & $2({\bf  1}, {\bf 10})(0,+1,\pm1,+2)_L$ &  $ ({\bf 1},{ {\bf 10}}) (\pm 1,-1,0,-2)_L$ \\
   & $2({\bf  1}, {\bf 1})(0,+1,\pm1,-4)_L$ &  $ ({\bf 1},{ {\bf 1}}) (\pm 1,-1,0,+4)_L$ \\
 & & \\
 \hline
  & & \\
  $T6$  &  $ ({\bf 1},{\overline {\bf 16}}) (\pm 1,-1,0,+1)_L$ & 2({\bf  1}, 
 ${\bf 16})(0,+1,\pm1,-1)_L$  \\
      &  $ ({\bf 1},{{\bf 10}}) (\pm 1,-1,0,-2)_L$ & 2({\bf  1}, 
 ${\bf 10})(0,+1,\pm1,+2)_L$  \\
     &  $ ({\bf 1},{{\bf 1}}) (\pm 1,-1,0,+4)_L$ & 2({\bf  1}, 
 ${\bf 1})(0,+1,\pm1,-4)_L$  \\

    & &  \\
 \hline
 & &  \\
   $T2$ & $({\bf 2},{\bf 1})(0,0,{\pm 3},0)_L$ & $2 ({\bf 1},{\bf 1})(0,-{3},\pm 3,0)_L$ \\
    &  $ ({\bf 1},{\bf 1})(\pm {3},+{3},0,0)_L$ & \\
  & &\\
   \hline
 & &  \\
 $U(1)$ & $(1/ \sqrt{6}, ~1/3\sqrt{2}, ~1/\sqrt{6},~1/6)$ & $(1/ \sqrt{6}, 
  ~1/{3\sqrt{2}},    ~1/\sqrt{6},~1/6)$\\
\end{tabular}
\caption{The massless spectra of the two $SO(10)$ models $T1(1,1)$ and $T2(1,1)$ both with gauge symmetry $SU(2)_1 \otimes SO(10)_3\otimes U(1)^4$.
The $U(1)$ normalization
radii are given at the bottom of the Table.
The gravity, dilaton and gauge supermultiplets are not shown.}

\end{table}

\begin{table}[t]
\begin{tabular}{|c|l|l|l|} 
 & $T1(0,0)$ & $T2(0,0)$ & $T3(0,0)$ \\
 M & $SU(2) \otimes SO(10) \otimes U(1)^4$ & 
      $SU(2) \otimes SO(10) \otimes U(1)^3$ &
  $SU(2) \otimes SO(10) \otimes U(1)^2$ \\ \hline
 & & & \\
   & $ ({\bf 1},{\bf 45})(0,0,0,0)_L$ 
    & $ ({\bf 1},{\bf 45})(0,0,0)_L$
    & $ ({\bf 1},{\bf 45})(0,0)_L$ \\
   & $2 ({\bf 1},{\bf 1})(0,+12,0,0)_L$ 
    & $2 ({\bf 1},{\bf 1})(0,+12,0)_L$
    & $4 ({\bf 1},{\bf 1})(0,0)_L$ \\
 $U$ & $2 ({\bf 1},{\bf 1})(0,0,+12,0)_L$ 
    &  $2 ({\bf 1},{\bf 1})(0,0,0)_L$ & \\
   & $3 ({\bf 1},{\bf 1})(0,-6,0,0)_L$ 
    &  $3 ({\bf 1},{\bf 1})(0,-6,0)_L$ & \\
   & $3 ({\bf 1},{\bf 1})(0,0,-6,0)_L$ & & \\ 
 & & & \\
\hline
 & & & \\
   &  $2 ({\bf 1},{\bf 16})(0,+{2},+{2},-1)_L$ 
    & $3 ({\bf 1},{\bf 16})(0,+{2},-1)_L$ 
     & $5 ({\bf 1},{\bf 16})(0,-1)_L$  \\
   & $2 ({\bf 1},{\bf 10})(0,+2,+2,+{2})_L$ 
    & $3 ({\bf 1},{\bf 10})(0,+2,+{2})_L$ 
    & $5 ({\bf 1},{\bf 10})(0,+{2})_L$ \\
   & $2 ({\bf 1},{\bf 1})(0,+{2},+{2},-4)_L$ 
    & $3 ({\bf 1},{\bf 1})(0,+{2},-4)_L$
    & $5 ({\bf 1},{\bf 1})(0,-4)_L$ \\
   & $ ({\bf 1},{\bf 16})(0,-{4},-{4},-{1})_L$ 
    & $2 ({\bf 1},{\bf 16})(0,-{4},-{1})_L$ & \\
 $T3$ & $ ({\bf 1},{\bf 10})(0,-{4},-{4},+{2})_L$ 
    & $ 2 ({\bf 1},{\bf 10})(0,-{4},+{2})_L$ & \\
   & $ ({\bf 1},{\bf 1})(0,-{4},-{4},-4)_L$ 
   &  $ 2 ({\bf 1},{\bf 1})(0,-{4},-4)_L$ & \\
   & $ ({\bf 1},{\bf 16})(0,-{4},+{2},-{1})_L$ & &\\
   & $ ({\bf 1},{\bf 10})(0,-{4},+{2},+{2})_L$ & & \\
   & $ ({\bf 1},{\bf 1})(0,-{4},+2,-{4})_L$ & & \\
   & $ ({\bf 1},{\bf 16})(0,+{2},-{4},-{1})_L$ & & \\
   & $ ({\bf 1},{\bf 10})(0,+{2},-{4},+{2})_L$ & & \\
   & $ ({\bf 1},{\bf 1})(0,+{2},-{4},-{4})_L$  & & \\
 & & & \\
 \hline
  & & & \\
   &  $ ({\bf 1},{\overline {\bf 16}}) (\pm 1,+{1},+{1},+{1})_L$ 
    & $ ({\bf 1},{\overline {\bf 16}}) (\pm 1,+{1},+{1})_L$
    & $ ({\bf 1},{\overline {\bf 16}}) (\pm 1,+{1})_L$ \\
 $T6$ &  $ ({\bf 1},{\bf 10})(\pm 1,+{1},+{1},-{2})_L$ 
     &  $ ({\bf 1},{\bf 10})(\pm 1,+{1},-{2})_L$ 
      &  $ ({\bf 1},{\bf 10})(\pm 1,-{2})_L$ \\
   & $ ({\bf 1},{\bf 1})(\pm 1,+{1},+{1},+{4})_L$ 
    & $ ({\bf 1},{\bf 1})(\pm 1,+{1},+{4})_L$ 
    & $ ({\bf 1},{\bf 1})(\pm 1,+{4})_L$ \\
 & & & \\
 \hline
 & & & \\
   &   $2 ({\bf 2},{\bf 1})(0,-{3},-{3},0)_L$ 
    & $2 ({\bf 2},{\bf 1})(0,{\pm 3},0)_L$ 
    & $6 ({\bf 2},{\bf 1})(0,0)_L$ \\
 $T2$ &  $  ({\bf 2},{\bf 1})(0,\pm {9},+{3},0)_L$ 
    & $ ({\bf 2},{\bf 1})(0,\pm {9},0)_L$ & \\
   &  $  ({\bf 2},{\bf 1})(0,+{3},\pm {9},0)_L$ & & \\
   &  $ ({\bf 1},{\bf 1})(\pm 3,-{3},-{3},0)_L$ 
    &  $ ({\bf 1},{\bf 1})(\pm 3,-{3},0)_L$
    &  $ ({\bf 1},{\bf 1})(\pm 3,0)_L$ \\ 
 & & & \\
\hline
 & & & \\
 $U(1)$ & $(1/ \sqrt{6},~1/{6\sqrt{2}}, ~1/{6\sqrt{2}},~1/6)$
   & $(1/ \sqrt{6},~1/{6\sqrt{2}},~1/6)$
   & $(1/ \sqrt{6},~1/6)$ \\
\end{tabular}
\caption{The massless spectra of the three models $T1(0,0)$, $T2(0,0)$ and
$T3(0,0)$ with gauge groups $SU(2)_1 \otimes SO(10)_3\otimes U(1)^4$,
$SU(2)_1 \otimes SO(10)_3 \otimes U(1)^3$, 
and $SU(2)_1 \otimes SO(10)_3 \otimes U(1)^2$, respectively. The $U(1)$ normalization
radii are given at the bottom of the Table. 
The gravity, dilaton and gauge supermultiplets are not shown.}

\end{table}

\begin{table}[t]
\begin{tabular}{|c||l|l|} 
 & $T4(0,0)$ &  $T5(0,0)$\\ 
M & $SU(2) \otimes SO(10)\otimes U(1)^3$ &
  $SU(2) \otimes SO(10) \otimes U(1)^2$ \\  \hline
 & & \\
   & $({\bf 1},{\bf 45})(0,0,0)_L$ & $({\bf 1},{\bf 45})(0,0)_L$ \\
   & $1({\bf 1},{\bf 1})(0,0,0)_L$  &  $2({\bf 1},{\bf 1})(0,0)_L$ \\
 $U$ & $ ({\bf 1},{\bf 1})(0,\pm6,0)_L$ & \\
    &  $({\bf 1},{\bf 1})(0,+12,0)_L$ & \\
  &  $({\bf 2},{\bf 1})(\pm 3,0,0)_L$ &
       $({\bf 2},{\bf 1})(\pm 3,0)_L$ \\
     & &  \\ \hline
 & &  \\
     & $ ({\bf 1},{\overline {\bf 16}}) (0,-4,+{1})_L$ & \\
  &$ ({\bf 1},{\bf 10})(0,-4,-{2})_L$ & \\
  &  $({\bf 1},{\bf 1})(0,-4,+{4})_L$ & \\
  & $ ({\bf 1},{\overline {\bf 16}}) (0,+2,+{1})_L$ &
      $2 ({\bf 1},{\overline {\bf 16}}) ( 0,+{1})_L$ \\
 $T3$  &$ ({\bf 1},{\bf 10})(0,+2,-{2})_L$ &
      $ 2({\bf 1},{\bf 10})( 0,-{2})_L$ \\
   &  $({\bf 1},{\bf 1})(0,+2,+{4})_L$ &
      $ 2({\bf 1},{\bf 1})( 0,+{4})_L$ \\
      & $ ({\bf 1},{\bf 16})(0,-{2},-{1})_L$ &
     $ ({\bf 1},{\bf 16})(0,-{1})_L$ \\
   & $ ({\bf 1},{\bf 10})(0,-{2},+{2})_L$ &
     $ ({\bf 1},{\bf 10})(0,+{2})_L$ \\
   & $({\bf 1},{\bf 1})(0,-{2},-{4})_L$ &
      $({\bf 1},{\bf 1})(0,-{4})_L$ \\
 & & \\  \hline
 & & \\
  & $2 ({\bf 1},{\bf 16})(\pm 1,+{1},-{1})_L$ &
     $2 ({\bf 1},{\bf 16})({\pm 1},-{1})_L$ \\
   $T6$ & $2 ({\bf 1},{\bf 10})(\pm 1,+{1},+2)_L$ &
     $2 ({\bf 1},{\bf 10})({\pm 1},+2)_L$ \\
   & $2 ({\bf 1},{\bf 1})(\pm 1,+{1},-{4})_L$ &
     $2 ({\bf 1},{\bf 1},{\bf 1})({\pm 1},-{4})_L$ \\
 & &  \\  \hline 
 & &  \\
  $T2$ & $2 ({\bf 1},{\bf 1})(\pm 3,-3,0)_L$ & $2 ({\bf 1},{\bf 1})(\pm 3,0)_L$ \\
   & $2 ({\bf 2},{\bf 1})(0,\pm 3,0)_L$  & $4 ({\bf 2},{\bf 1})(0,0)_L$\\
    & & \\ \hline
& & \\
 $U(1)$ & $(1/ \sqrt{6},~1/{6\sqrt{2}},~1/6)$ &
    $(1/ \sqrt{6},~1/6)$ \\
\end{tabular}
\caption{The massless spectra of the $T4(0,0)$ and $T5(0,0)$ models 
with gauge groups $SU(2)_1\otimes SO(10)_3\otimes U(1)^3$
and $SU(2)_1\otimes SO(10)_3\otimes U(1)^2$, respectively. 
The $U(1)$ normalization radii are given at the bottom of the Table.
The graviton, dilaton and gauge supermultiplets are not shown.}
\end{table}

\begin{table}[t]
\begin{tabular}{|c|l|l|l|} 
 & $T9(0,0)$ & $T10(0,0)$ & $T11(0,0)$\\
 M & $SU(2) \otimes SO(10) \otimes U(1)^4$ &  $SU(2) \otimes SO(10) \otimes U(1)^3$ 
  & $SU(2) \otimes SO(10) \otimes U(1)^2$ \\ \hline
 & & & \\
   & $ ({\bf 1},{\bf 45})(0,0,0,0)_L$ & $ ({\bf 1},{\bf 45})(0,0,0)_L$ & 
   $ ({\bf 1},{\bf 45})(0,0)_L$\\
      & $ ({\bf 1},{\bf 1})(0,+12,0,0)_L$  & $ ({\bf 1},{\bf 1})(0,+12,0)_L$
   & $2 ({\bf 1},{\bf 1})(0,0)_L$\\
 $U$ & $ ({\bf 1},{\bf 1})(0,0,+12,0)_L$  & $({\bf 1},{\bf 1})(0,\pm 6,0)_L$ & \\
   & $({\bf 1},{\bf 1})(0,\pm 6,0,0)_L$  & $({\bf 1},{\bf 1})(0,0,0)_L$ & \\
   & $({\bf 1},{\bf 1})(0,0,\pm6,0)_L$ & & \\ 
 & & &\\
\hline
 & & & \\
     & $ ({\bf 1},{\overline {\bf 16}})(0,-{4},-{4},+{1})_L$  
    & $ ({\bf 1},{\overline {\bf 16}})(0,-{4},+{1})_L$
    & $ 2({\bf 1},{\overline {\bf 16}})(0,+{1})_L$\\
     & $ ({\bf 1},{\bf 10})(0,-{4},-{4},-{2})_L$ 
     & $ ({\bf 1},{\bf 10})(0,-{4},-{2})_L$
     & $ 2({\bf 1},{\bf 10})(0,-{2})_L$\\
     & $ ({\bf 1},{\bf 1})(0,-{4},-{4},+{4})_L$ 
     & $ ({\bf 1},{\bf 1})(0,-{4},+{4})_L$
     & $ 2({\bf 1},{\bf 1})(0,+{4})_L$  \\
 $T3$  & $ ({\bf 1},{\overline {\bf 16}})(0,+2,-4,+{1})_L$
                 & $ ({\bf 1},{\overline {\bf 16}})(0,+2,+{1})_L$  & \\
   & $ ({\bf 1},{\bf 10})(0,+2,-4,-2)_L$ 
    & $ ({\bf 1},{\bf 10})(0,+2,-2)_L$  & \\
   & $ ({\bf 1},{\bf 1})(0,+2,-4,+{4})_L$ 
    & $ ({\bf 1},{\bf 1})(0,+2,+{4})_L$  & \\
   & $ ({\bf 1},{\bf 16})(0,-{2},+{4},-{1})_L$ 
   & $ ({\bf 1},{\bf 16})(0,-{2},-{1})_L$
   & $ ({\bf 1},{\bf 16})(0,-{1})_L$ \\
   & $ ({\bf 1},{\bf 10})(0,-{2},+{4},+{2})_L$
   & $ ({\bf 1},{\bf 10})(0,-{2},+{2})_L$
    & $ ({\bf 1},{\bf 10})(0,+{2})_L$\\
   & $ ({\bf 1},{\bf 1})(0,-{2},+{4},-{4})_L$ 
    & $ ({\bf 1},{\bf 1})(0,-{2},-{4})_L$
    & $ ({\bf 1},{\bf 1})(0,-{4})_L$\\
 & & & \\
 \hline
  & & &\\
   &  $2 ({\bf 1},{{\bf 16}}) (\pm 1,+{1},+{1},-{1})_L$  
    & $2 ({\bf 1},{{\bf 16}}) (\pm 1,+{1},-{1})_L$
    & $2 ({\bf 1},{{\bf 16}}) (\pm 1,-{1})_L$\\
 $T6$ 
   &  $2 ({\bf 1},{\bf 10})(\pm 1,+{1},+{1},+{2})_L$ 
    &  $2 ({\bf 1},{\bf 10})(\pm 1,+{1},+{2})_L$
    &  $2 ({\bf 1},{\bf 10})(\pm 1,+{2})_L$ \\
   & $2 ({\bf 1},{\bf 1})(\pm 1,+{1},+{1},-{4})_L$
   & $2 ({\bf 1},{\bf 1})(\pm 1,+{1},-{4})_L$
    & $2 ({\bf 1},{\bf 1})(\pm 1,-{4})_L$ \\
 & & & \\
 \hline
 & & & \\
   &   $2 ({\bf 1},{\bf 1})(\pm 3,-{3},-{3},0)_L$ 
    &   $4 ({\bf 1},{\bf 1})(\pm 3,-{3},0)_L$
    &   $6 ({\bf 1},{\bf 1})(\pm 3,0)_L$\\
    & $2 ({\bf 1},{\bf 1})(\pm 3, +3,-3,0)_L$ 
    & $2 ({\bf 1},{\bf 1})(\pm 3, +3,0)_L$ & \\
    & $2 ({\bf 1},{\bf 1})(\pm 3, -3,+3,0)_L$  & & \\
 $T2$ 
     &  $ 2 ({\bf 2},{\bf 1})(0,\pm {9},+{3},0)_L$ 
     &  $ 2 ({\bf 2},{\bf 1})(0,\pm {9},0)_L$
     &  $ 20({\bf 2},{\bf 1})(0,0)_L$\\
    & $ 2({\bf 2},{\bf 1})(0,+3,\pm {9},0)_L$ 
    & $ 8({\bf 2},{\bf 1})(0,\pm 3,0)_L$ &\\
   &  $ 4 ({\bf 2},{\bf 1})(0,-3,-3,0)_L$ & & \\
   &  $ 4({\bf 2},{\bf 1})(0,\pm {3},\mp {3},0)_L$ & & \\
 & & &\\
\hline
 &  & & \\
 $U(1)$ & $(1/ \sqrt{6},~1/{6\sqrt{2}}, ~1/{6\sqrt{2}},~1/6)$ 
               & $(1/ \sqrt{6},~1/{6\sqrt{2}}, ~1/6)$
              & $(1/ \sqrt{6}, ~1/6)$\\
\end{tabular}
\caption{The massless spectra of the three models $T9(0,0)$, $T10(0,0)$ and 
$T11(0,0)$ with gauge groups
$SU(2)_1 \otimes SO(10)_3\otimes U(1)^4$,
$SU(2)_1 \otimes SO(10)_3\otimes U(1)^3$, and 
$SU(2)_1 \otimes SO(10)_3\otimes U(1)^2$, respectively. 
The $U(1)$ normalization radii are given at the bottom of the Table. 
The gravity, dilaton and gauge supermultiplets are not shown.}

\end{table}

\begin{table}[t]
\begin{tabular}{|c||l|l|} 
 & $T1(1,0)$ &  $T2(1,0)$\\ 
M & $SU(2)^2 \otimes SO(10)\otimes U(1)^3$ &
  $SU(2)^2 \otimes SO(10) \otimes U(1)^3$ \\  \hline
 & & \\
   & $({\bf 1},{\bf 1},{\bf 45})(0,0,0)_L$ & $({\bf 1},{\bf 1},{\bf 45})(0,0,0)_L$ \\
   & $({\bf 1},{\bf 3},{\bf 1})(0,0,0)_L$  &  $({\bf 1},{\bf 3},{\bf 1})(0,0,0)_L$ \\
 $U$ & $ ({\bf 1},{\bf 1},{\bf 1})(0,-6,0)_L$ & 
     $ ({\bf 1},{\bf 1},{\bf 1}) (0,+6,0)_L$  \\
   & $2 ({\bf 1},{\bf 4},{\bf 1})(0,+3,0)_L$ & 
     $2 ({\bf 1},{\bf 1},{\bf 1})({\pm 3},-3,0)_L$ \\
   & $2 ({\bf 1},{\bf 2},{\bf 1})(0,-3,0)_L$  & $2 ({\bf 2},{\bf 2},{\bf 1})(0,0,0)_L$\\
 & &  \\ \hline
 & &  \\
   & $2 ({\bf 1},{\bf 2},{\bf 16})(0,-{1},-{1})_L$ &
     $2 ({\bf 1},{\bf 1},{\bf 16})({\pm 1},+1,-{1})_L$ \\
   & $2 ({\bf 1},{\bf 2},{\bf 10})(0,-{1},+2)_L$ &
     $2 ({\bf 1},{\bf 1},{\bf 10})({\pm 1},+1,+2)_L$ \\
  $T3$ & $2 ({\bf 1},{\bf 2},{\bf 1})(0,-{1},-{4})_L$ &
     $2 ({\bf 1},{\bf 1},{\bf 1})({\pm 1},+1,-{4})_L$ \\
   & $ ({\bf 1},{\bf 1},{\bf 16})(0,+{2},-{1})_L$ &
     $ ({\bf 1},{\bf 1},{\bf 16})(0,-{2},-{1})_L$ \\
   & $ ({\bf 1},{\bf 1},{\bf 10})(0,+{2},+{2})_L$ &
     $ ({\bf 1},{\bf 1},{\bf 10})(0,-{2},+{2})_L$ \\
   & $({\bf 1},{\bf 1},{\bf 1})(0,+{2},-{4})_L$ &
      $({\bf 1},{\bf 1},{\bf 1})(0,-{2},-{4})_L$ \\
 & & \\  \hline
 & & \\
  & $ ({\bf 1},{\bf 1},{\overline {\bf 16}}) (\pm 1,+{1},+{1})_L$ &
      $ ({\bf 1},{\bf 2},{\overline {\bf 16}}) ( 0,-{1},+{1})_L$ \\
 $T6$  &$ ({\bf 1},{\bf 1},{\bf 10})(\pm 1,+{1},-{2})_L$ &
      $ ({\bf 1},{\bf 2},{\bf 10})( 0,-{1},-{2})_L$ \\
   &  $({\bf 1},{\bf 1},{\bf 1})(\pm 1,+{1},+{4})_L$ &
      $ ({\bf 1},{\bf 2},{\bf 1})( 0,-{1},+{4})_L$ \\
 & &  \\  \hline 
 & &  \\
   &  $({\bf 2},{\bf 2},{\bf 1})(0,0,0)_L$ &
      $({\bf 2},{\bf 1},{\bf 1})(\pm 3,0,0)_L$ \\
 $T2$ &  $({\bf 2},{\bf 4},{\bf 1})(0,0,0)_L$ &
       $({\bf 1},{\bf 4},{\bf 1})(0,+3,0)_L$ \\
   & $({\bf 1},{\bf 1},{\bf 1})(\pm 3,-3,0)_L$ &
   $ 2 ({\bf 1},{\bf 2},{\bf 1})(0,-3,0)_L$ \\
   &   & $({\bf 1},{\bf 2},{\bf 1})(0,+3,0)_L$  \\
 & & \\ \hline
& & \\
 $U(1)$ & $(1/ \sqrt{6},~1/{3\sqrt{2}},~1/6)$ &
    $(1/ \sqrt{6},~1/{3\sqrt{2}},~1/6)$ \\
\end{tabular}
\caption{The massless spectra of the $T1(1,0)$ and $T2(1,0)$ models 
both with gauge symmetry $SU(2)_1\otimes SU(2)_3 \otimes SO(10)_3\otimes U(1)^3$. The $U(1)$ normalization radii are given at the bottom of the Table.
The graviton, dilaton and gauge supermultiplets are not shown.}
\end{table}

\begin{table}[t]
\begin{tabular}{|c||l|l|} 
 & $T3(1,0)$ &  $T4(1,0)$\\ 
M & $SU(2)^2 \otimes SO(10)\otimes U(1)^3$ &
  $SU(2)^2 \otimes SO(10) \otimes U(1)^3$ \\  \hline
 & & \\
   & $({\bf 1},{\bf 1},{\bf 45})(0,0,0)_L$ & $({\bf 1},{\bf 1},{\bf 45})(0,0,0)_L$ \\
   & $({\bf 1},{\bf 3},{\bf 1})(0,0,0)_L$  &  $({\bf 1},{\bf 3},{\bf 1})(0,0,0)_L$ \\
 $U$ & $ ({\bf 1},{\bf 1},{\bf 1})(0,-6,0)_L$ & 
     $ ({\bf 1},{\bf 1},{\bf 1}) (0,+6,0)_L$  \\
    &  $({\bf 2},{\bf 2},{\bf 1})(0,0,0)_L$ &
      $({\bf 2},{\bf 1},{\bf 1})(\pm 3,0,0)_L$ \\
  &  $({\bf 2},{\bf 4},{\bf 1})(0,0,0)_L$ &
       $({\bf 1},{\bf 4},{\bf 1})(0,+3,0)_L$ \\
   & $({\bf 1},{\bf 1},{\bf 1})(\pm 3,-3,0)_L$ &
   $ 2 ({\bf 1},{\bf 2},{\bf 1})(0,-3,0)_L$ \\
   &   & $({\bf 1},{\bf 2},{\bf 1})(0,+3,0)_L$  \\
  & &  \\ \hline
 & &  \\
     & $ ({\bf 1},{\bf 1},{\overline {\bf 16}}) (\pm 1,+{1},+{1})_L$ &
      $ ({\bf 1},{\bf 2},{\overline {\bf 16}}) ( 0,-{1},+{1})_L$ \\
 $T3$  &$ ({\bf 1},{\bf 1},{\bf 10})(\pm 1,+{1},-{2})_L$ &
      $ ({\bf 1},{\bf 2},{\bf 10})( 0,-{1},-{2})_L$ \\
   &  $({\bf 1},{\bf 1},{\bf 1})(\pm 1,+{1},+{4})_L$ &
      $ ({\bf 1},{\bf 2},{\bf 1})( 0,-{1},+{4})_L$ \\
      & $ ({\bf 1},{\bf 1},{\bf 16})(0,+{2},-{1})_L$ &
     $ ({\bf 1},{\bf 1},{\bf 16})(0,-{2},-{1})_L$ \\
   & $ ({\bf 1},{\bf 1},{\bf 10})(0,+{2},+{2})_L$ &
     $ ({\bf 1},{\bf 1},{\bf 10})(0,-{2},+{2})_L$ \\
   & $({\bf 1},{\bf 1},{\bf 1})(0,+{2},-{4})_L$ &
      $({\bf 1},{\bf 1},{\bf 1})(0,-{2},-{4})_L$ \\
 & & \\  \hline
 & & \\
  & $2 ({\bf 1},{\bf 2},{\bf 16})(0,-{1},-{1})_L$ &
     $2 ({\bf 1},{\bf 1},{\bf 16})({\pm 1},+1,-{1})_L$ \\
   $T6$ & $2 ({\bf 1},{\bf 2},{\bf 10})(0,-{1},+2)_L$ &
     $2 ({\bf 1},{\bf 1},{\bf 10})({\pm 1},+1,+2)_L$ \\
   & $2 ({\bf 1},{\bf 2},{\bf 1})(0,-{1},-{4})_L$ &
     $2 ({\bf 1},{\bf 1},{\bf 1})({\pm 1},+1,-{4})_L$ \\
 & &  \\  \hline 
 & &  \\
  $T2$ & $2 ({\bf 1},{\bf 4},{\bf 1})(0,+3,0)_L$ & 
     $2 ({\bf 1},{\bf 1},{\bf 1})({\pm 3},-3,0)_L$ \\
   & $2 ({\bf 1},{\bf 2},{\bf 1})(0,-3,0)_L$  & $2 ({\bf 2},{\bf 2},{\bf 1})(0,0,0)_L$\\
    & & \\ \hline
& & \\
 $U(1)$ & $(1/ \sqrt{6},~1/{3\sqrt{2}},~1/6)$ &
    $(1/ \sqrt{6},~1/{3\sqrt{2}},~1/6)$ \\
\end{tabular}
\caption{The massless spectra of the $T3(1,0)$ and $T4(1,0)$ models 
both with gauge symmetry $SU(2)_1\otimes SU(2)_3 \otimes SO(10)_3\otimes U(1)^3$. The $U(1)$ normalization radii are given at the bottom of the Table.
The graviton, dilaton and gauge supermultiplets are not shown.}
\end{table}

\begin{table}[t]
\begin{tabular}{|c||l|l|} 
 & $T5(1,0)$ &  $T6(1,0)$\\ 
M & $SU(2)^3 \otimes SO(10)\otimes U(1)^4$ &
  $SU(2)^3 \otimes SO(10) \otimes U(1)^4$ \\  \hline
   & & \\
   & $({\bf 1},{\bf 1},{\bf 1},{\bf 45})(0,0,0,0)_L$ & $({\bf 1},{\bf 1},{\bf 1},{\bf 45})(0,0,0,0)_L$ \\
   & $ ({\bf 1},{\bf 1},{\bf 1},{\bf 1})(\pm 3,+3,0,0)_L$  
   &  $({\bf 2},{\bf 1},{\bf 1},{\bf 1})(\pm 3,0,0,0)_L$ \\
 $U$ 
   &  $({\bf 1},{\bf 1},{\bf 1},{\bf 1})(0,+6,0,0)_L$ & $ ({\bf 1},{\bf 1},{\bf 1},{\bf 1}) (0,+6,0,0)_L$  \\
   &  $({\bf 1},{\bf 1},{\bf 1},{\bf 1})(0,0,+6,0)_L$ & $({\bf 1},{\bf 1},{\bf 1},{\bf 1})(0,0,+6,0)_L$ \\
   &  $ ({\bf 2},{\bf 2},{\bf 2},{\bf 1})(0,0,+3,0)_L$ & $({\bf 1},{\bf 2},{\bf 2},{\bf 1})(0,-3,-3,0)_L$ \\
   &  & $({\bf 1},{\bf 2},{\bf 2},{\bf 1})(0,\pm 3,\mp 3,0)_L$ \\
   &  &  \\ \hline
     & &  \\
     & $ ({\bf 1},{\bf 1},{\bf 1},{\overline {\bf 16}}) ( 0,+2,+2,+1)_L$
     & $ ({\bf 1},{\bf 1},{\bf 1},{\overline {\bf 16}}) ( 0,+2,+2,+1)_L$ \\
 $T3$  
     & $ ({\bf 1},{\bf 1}, {\bf 1},{\bf 10})( 0,+2,+{2},-2)_L$
     & $ ({\bf 1},{\bf 1}, {\bf 1},{\bf 10})( 0,+2,+{2},-2)_L$ \\
     & $ ({\bf 1},{\bf 1},{\bf 1},{\bf 1})( 0,+2,+{2},+4)_L$
     & $ ({\bf 1},{\bf 1},{\bf 1},{\bf 1})( 0,+2,+{2},+4)_L$ \\
       & $({\bf 1},{\bf 1},{\bf 1},{\bf 16})({\pm 1},+1,-{2},-1)_L$  & \\
       & $({\bf 1},{\bf 1},{\bf 1},{\bf 10})({\pm 1},+1,-2,+2)_L$  & \\
       & $ ({\bf 1},{\bf 1},{\bf 1},{\bf 1})({\pm 1},+1,-{2},-4)_L$ & \\
      &  & \\  \hline
 & & \\
     &  & $({\bf 1},{\bf 1},{\bf 1},{\bf 16})({\pm 1},+1,-{2},-1)_L$ \\
   $T6$ 
     &  & $({\bf 1},{\bf 1},{\bf 1},{\bf 10})({\pm 1},+1,-2,+2)_L$ \\
     &  & $ ({\bf 1},{\bf 1},{\bf 1},{\bf 1})({\pm 1},+1,-{2},-4)_L$ \\
    &  $({\bf 1},{\bf 1},{\bf 1},{\bf 16})({\pm 1},-{2},+1,-1)_L$
    & $({\bf 1},{\bf 1},{\bf 1},{\bf 16})({\pm 1},-{2},+1,-1)_L$ \\
     & $({\bf 1},{\bf 1},{\bf 1},{\bf 10})({\pm 1},-2,+1+2)_L$
     & $({\bf 1},{\bf 1},{\bf 1},{\bf 10})({\pm 1},-2,+1+2)_L$ \\
     &  $ ({\bf 1},{\bf 1},{\bf 1},{\bf 1})({\pm 1},-{2},+1,-4)_L$ 
     & $ ({\bf 1},{\bf 1},{\bf 1},{\bf 1})({\pm 1},-{2},+1,-4)_L$ \\
     &  &  \\  \hline 
 & &  \\
  $T2$     & $ ({\bf 1},{\bf 2},{\bf 1},{\bf 1})(\pm 3,0,0,0)_L$
                & $ ({\bf 1},{\bf 2},{\bf 1},{\bf 1})(\pm 3,0,0,0)_L$ \\
              &  $ ({\bf 1},{\bf 1},{\bf 2},{\bf 1})(\pm 3,0,0,0)_L$
              & $ ({\bf 1},{\bf 1},{\bf 2},{\bf 1})(\pm 3,0,0,0)_L$ \\
               & $({\bf 2},{\bf 1},{\bf 1},{\bf 1})(\pm 3,0,0,0)_L$ 
               & $ ({\bf 1},{\bf 1},{\bf 1},{\bf 1})(\pm 3,+3,0,0)_L$ \\
               &  $ ({\bf 1},{\bf 1},{\bf 1},{\bf 1})(\pm 3,0,+3,0)_L$ 
               & $ ({\bf 1},{\bf 1},{\bf 1},{\bf 1})(\pm 3,0,+3,0)_L$ \\
            & $ ({\bf 2},{\bf 1},{\bf 2},{\bf 1})(0,-3,-3,0)_L$ 
            & $ ({\bf 2},{\bf 1},{\bf 2},{\bf 1})(0,-3,-3,0)_L$ \\
            &  $ ({\bf 2},{\bf 1},{\bf 2},{\bf 1})(0,\pm 3,\mp 3,0)_L$
            & $ ({\bf 2},{\bf 1},{\bf 2},{\bf 1})(0,\pm 3,\mp 3,0)_L$ \\
            &   $ ({\bf 2},{\bf 2},{\bf 1},{\bf 1})(0,-3,-3,0)_L$ 
            & $ ({\bf 2},{\bf 2},{\bf 1},{\bf 1})(0,-3,-3,0)_L$ \\
            &  $ ({\bf 2},{\bf 2},{\bf 1},{\bf 1})(0,\pm 3,\mp 3,0)_L$
            & $ ({\bf 2},{\bf 2},{\bf 1},{\bf 1})(0,\pm 3,\mp 3,0)_L$ \\
            &  $ ({\bf 2},{\bf 2},{\bf 2},{\bf 1})(0,+3,0,0)_L$
            & $ ({\bf 2},{\bf 2},{\bf 2},{\bf 1})(0,+3,0,0)_L$ \\
              &  $({\bf 1},{\bf 2},{\bf 2},{\bf 1})(0,-3,-3,0)_L$ 
              & $ ({\bf 2},{\bf 2},{\bf 2},{\bf 1})(0,0,+3,0)_L$ \\
              & $({\bf 1},{\bf 2},{\bf 2},{\bf 1})(0,\pm 3,\mp 3,0)_L$ & \\
    & & \\ \hline
 $U(1)$ &  $(1/ \sqrt{6},~1/{3\sqrt{2}},~1/{3\sqrt{2}},~1/6)$ 
              & $(1/ \sqrt{6},~1/{3\sqrt{2}},~1/{3\sqrt{2}},~1/6)$ \\
\end{tabular}
\caption{The massless spectra of the $T5(1,0)$ and $T6(1,0)$ models 
both with gauge symmetry $SU(2)^3_1\otimes SO(10)_3\otimes U(1)^4$. The $U(1)$ normalization radii are given at the bottom of the Table.
The graviton, dilaton and gauge supermultiplets are not shown.}
\end{table}

\begin{table}[t]
\begin{tabular}{|c||l|l|l|} 
  & $T6(0,0)$ &  $T7(0,0)$ & $T8(0,0)$ \\
 & $SU(2)_1 \otimes SO(10)_3\otimes U(1)^4$
 & $SU(2)_1 \otimes SO(10)_3\otimes U(1)^3$
 & $SU(2)_1 \otimes SO(10)_3\otimes U(1)^2$ \\
\hline
   & & & \\
   & $2 ({\bf 2},{\bf 1})(0,\pm 3,\pm 3,0)_L$ 
   &  $6 ({\bf 2},{\bf 1})(0,\pm 3,0)_L$ 
   &   $16 ({\bf 2},{\bf 1})(0,0)_L$\\
   & $2 ({\bf 2},{\bf 1})(0,\pm 3,\mp 3,0)_L$ & &  \\
   & $ ({\bf 2},{\bf 1})(0,+3,\pm 9,0)_L$ & &  \\
 $T2$ & $ ({\bf 2},{\bf 1})(0,-3,\pm 9,0)_L$ & &\\
    & $ ({\bf 2},{\bf 1})(0,\pm 9,+3,0)_L$ & $ 2({\bf 2},{\bf 1})(0,\pm 9,0)_L$ & \\
    & $ ({\bf 2},{\bf 1})(0,\pm 9,-3,0)_L$ &  & \\
    & $ ({\bf 1},{\bf 1})(\pm 3,+3,+3,0)_L$ 
    & $ 2 ({\bf 1},{\bf 1})(\pm 3,+3,0)_L$ 
     & $ 4({\bf 1},{\bf 1})(\pm 3,0)_L$\\
    & $ ({\bf 1},{\bf 1})(\pm 3,+3,-3,0)_L$ & $ 2({\bf 1},{\bf 1})(\pm 3,-3,0)_L$& \\
    & $ ({\bf 1},{\bf 1})(\pm 3,-3,+3,0)_L$ & &\\
    & $ ({\bf 1},{\bf 1})(\pm 3,-3,-3,0)_L$ & &\\
 &  & &\\ \hline
&  & &\\
 $U(1)$ 
  & $(1/ \sqrt{6},~1/{6\sqrt{2}},~1/{6\sqrt{2}},~1/6)$  
   & $(1/ \sqrt{6},~1/{6\sqrt{2}},~1/6)$
   & $(1/ \sqrt{6},~1/6)$
\end{tabular}
\caption{The states to be added to the massless spectra of the models 
$T1(0,0)$, $T2(0,0)$ and $T3(0,0)$ to obtain the massless spectra of 
the models $T6(0,0)$, $T7(0,0)$ and $T8(0,0)$, respectively. 
The $U(1)$ normalization radii are given at the bottom of the Table. 
The additional states shown in the Table appear in the $T2$ sector.}
\end{table}

\begin{table}[t]
\begin{tabular}{|c||l|} 
  & $T7(1,0)$, $T8(1,0)$, $T9(1,0)$, $T10(1,0)$ \\
  & $SU(2)_1 \otimes SU(2)_3 \otimes SO(10)_3 \otimes U(1)^3$\\
\hline
   &\\
    & $2 ({\bf 1},{\bf 2},{\bf 1})(\pm 3,0,0)_L$  \\
   $T2$ & $2 ({\bf 2},{\bf 1},{\bf 1})(0,\pm3,0)_L$  \\
    & $2 ({\bf 2},{\bf 3},{\bf 1})(0,\pm3,0)_L$  \\
  & \\ \hline
 &\\
 $U(1)$  & $(1/ \sqrt{6},~1/{3\sqrt{2}},~1/6)$ 
\end{tabular}
\caption{The states to be added to the massless spectra of the 
models $T1(1,0)$, $T2(1,0)$, $T3(1,0)$ and $T4(1,0)$ to obtain the massless 
spectra of the models $T7(1,0)$, $T8(1,0)$, $T9(1,0)$ and $T10(1,0)$, 
respectively. All of these models have gauge symmetry 
$SU(2)_1 \otimes SU(2)_3 \otimes SO(10)_3 \otimes U(1)^3$. 
The $U(1)$ normalization radii are given at the bottom of the Table. 
The additional states shown in the Table appear in the $T2$ sector.}
\end{table}


\end{document}